\documentclass{aa}  
\pdfoutput=1
\usepackage{graphicx}
\usepackage{txfonts}
\usepackage{bm}
\usepackage{natbib,twoopt}
\bibpunct{(}{)}{;}{a}{}{,}
\usepackage{color}

\begin{document}

   \title {Detecting and quantifying stellar magnetic fields}
   
   \subtitle{Sparse Stokes profile approximation using orthogonal matching pursuit}

   \author{T.A. Carroll
          \and    
          K.G. Strassmeier        
          }
   \offprints{T. A. Carroll}
   
   \institute{Leibniz-Institut f\"ur Astrophysik (AIP),
              An der Sternwarte 16, D-14482 Potsdam, Germany \\
              \email{tcarroll@aip.de,kstrassmeier@aip.de}
             }

   \date{Received October 10, 2013; accepted October 11, 2013}

 
  \abstract
   {In recent years, we have seen a rapidly growing number of stellar magnetic field 
   detections for various types of stars. 
   Many of these magnetic fields are estimated from spectropolarimetric observations (Stokes~$V$) by using the so-called 
   center-of-gravity (COG) method.  
   Unfortunately, the accuracy of this method rapidly deteriorates with increasing 
   noise and thus calls for a more robust procedure that combines signal detection and field estimation.  
   }
   {We introduce an estimation method that provides not only the effective or mean longitudinal magnetic
   field from an observed Stokes~$V$ profile but also uses the net absolute polarization 
   of the profile to obtain an estimate of the apparent (i.e., velocity resolved) absolute 
   longitudinal magnetic field. 
   }
   {
   By combining the COG method
   with an orthogonal-matching-pursuit (OMP) approach, we were able to decompose observed Stokes profiles
   with an overcomplete dictionary of wavelet-basis functions to reliably reconstruct the 
   observed Stokes profiles in the presence of noise.
   The elementary wave functions of the sparse reconstruction process were utilized to
   estimate the effective longitudinal magnetic field and the 
   apparent absolute longitudinal magnetic field. A multiresolution analysis 
   complements the OMP algorithm to provide a robust detection and estimation method. 
   }
   {   
   An extensive Monte-Carlo simulation confirms the reliability and accuracy 
   of the magnetic OMP approach where a mean error of under 2\% is found. 
   Its full potential is obtained for heavily noise-corrupted Stokes profiles 
   with signal-to-noise variance ratios down to unity. In this case a conventional 
   COG method yields a mean error for the effective longitudinal magnetic field of up to 50\%, 
   whereas the OMP method gives a maximum error of 18\%. 
   It is, moreover, shown that even in the case of very small residual noise on a level between 10$^{-3}$ and 10$^{-5}$, 
   a regime reached by current multiline reconstruction techniques, the conventional COG method 
   incorrectly interprets a large portion of the residual noise as a magnetic field, with values of up to 100 G. 
   The magnetic OMP method, on the other hand, remains largely unaffected by the noise, regardless of the noise 
   level the maximum error is no greater than 0.7 G.     
   }
   {}


   \maketitle
%

\section{Introduction}
\label{Sect:1}
There are various ways of obtaining information about the
presence of magnetic fields in stellar atmospheres like 
Ca II H \& K or X-ray measurements. But the most direct
way is certainly provided via the Zeeman effect, i.e., the magnetically induced 
line splitting and the associated line polarization. This has opened the way for 
essentially two different approaches to measuring the magnetic fields of cool
stars. Unpolarized Zeeman broadening measurements from Stokes~$I$ profiles has led to a 
large amount of information about the magnetic fields of cool stars 
over the past decades \citep[e.g.,][]{Robinson80,Saar88,Valenti95,Johns-Krull96,Johns-Krull07}. 
Because the Zeeman splitting (and broadening) 
is directly proportional to the magnetic field strength, Zeeman broadening
measurements provide good estimates of the underlying average surface field.
In the stellar astrophysical literature, the term magnetic flux for
the product of surface filling factor and magnetic field strength ($f \cdot B$) is often 
used, but we prefer to denote it more accurately as the mean or average 
magnetic field strength, since  $f \cdot B$ is neither the (relative) magnetic flux through the 
stellar surface nor a measure of the magnetic flux density.
Polarization or spectropolarimetric measurements, on the other hand, provide additional information 
about the orientation of the field vector and therefore allow, in principle, measurement of real flux densities. 

There are at least two obstacles that may prevent the 
full applicability of spectropolarimetric measurements to cool stars. First, the magnetic fields
are relatively small in size and magnitude, which leads to very weak circular polarization (Stokes~$V$)
and even lower linear polarization (Stokes~Q and U) signals. Second, the contributions 
from different polarities tend to mutually cancel out the circular polarization signal. 
A circumstance that mitigates the situation to some extent is rapid rotation, which partly lifts the
mutual cancellation of the Stokes~$V$ spectra. 
This has led to the so-called Zeeman-Doppler imaging
(ZDI) or Magnetic-Doppler imaging (MDI) methods \citep{Semel89,Donati97,Piskunov02}, 
which allows reconstructing, in an inversion approach, the entire surface distribution 
of the magnetic field vector from phase-resolved spectropolarimetric observations.
This approach has contributed significantly to our understanding 
of cool stars magnetic fields and stellar activity in general in recent years 
\citep[e.g.,][]{Donati03,Koch04,Donati06,Petit08,Donati08,Hussain09,Morin10,Carroll12}.
However, what is good for the ZDI approach, namely rapid rotation, has an adverse effect on the 
Zeeman broadening approach and vice versa, such that both approaches are 
complementary to some degree; see also \citet{Reiners12} for a comprehensive overview 
of measuring stellar magnetic fields on cool stars.

The present paper focuses on spectropolarimetric observations that
do not deliver a series of phase-resolved spectra but rather single 
snapshot-like observations from either rapidly or slowly rotating 
active stars. This is the typical situation for stars with long rotation 
periods and/or in situations where the amount of telescope time is limited.
Such observations are responsible for a number of exciting magnetic field detection in 
recent times \citep[e.g.,][]{Auriere07,Auriere09,Ligni09,Auriere10,Grunhut10,Konsta10,Sennhauser11,Petit11,Fossati13}.
Because many of the observed polarimetric line profiles are buried in noise, the investigations rely on a prior 
line profile reconstruction technique, such as the least-square-deconvolution (LSD) 
\citep{Donati97}, principal-component-analysis (PCA) \citep{Carroll07a}, or singular-value-decomposition (SVD) \citep{Carroll12}.
However, even after these preprocessing steps, the extracted line profiles exhibit a fraction of noise that makes
their interpretation not always straightforward in terms of a reliable detection and magnetic field
estimation.

Inspired by the work of \citet{Asensio10} who propose a compressive sensing framework for spectropolarimetry, 
we utilize the compressibility of polarimetric signals to approximate the observed Stokes~$V$ profiles by a sparse
decomposition with a wavelet frame. This sparse linear representation facilitates a 
simple and noise-robust magnetic detection and estimation method.\footnote{An IDL-Code of the magnetic OMP method
is available under http://www.aip.de/People/tcarroll and http://www.aip.de/Members/tcarroll}
For rapidly rotating stars, the sparse approximation of the observed signal with basis functions allows not only 
the effective mean longitudinal magnetic field to be measured but also the absolute value 
of the resolved (i.e. apparent) longitudinal magnetic field.

The paper is organized as follows. In Sect. \ref{Sect:2} we describe the weak-field approximation
and derive the center-of-gravity method for disk-integrated observations to 
define the effective longitudinal magnetic field. Furthermore,
we also define the apparent longitudinal magnetic field that describes the disk-integrated 
absolute value of the longitudinal magnetic field component 
as measured from the net absolute circular polarization of the Stokes~$V$ profile. 
In Sect. \ref{Sect:3} we introduce
the concept of a sparse Stokes profile approximation and magnetic field detection. 
We describe in some detail the orthogonal matching pursuit (OMP) algorithm and the signal dictionary, 
which provides the building blocks for the Stokes profile approximation. 
At the end of this section, we combine the approximation method with a detection algorithm to obtain 
reliable estimates for the magnetic quantities. 
Section \ref{Sect:4} gives an illustration of the magnetic OMP method
and highlights the benefit of the complementary definition of the effective and apparent longitudinal magnetic field.
In Sect.\ref{Sect:5} we investigate the sparsity of Stokes profiles and demonstrate that with the described
dictionary a sparse approximation of the Stokes profile is possible for a wide parameter range.
An extensive numerical assessment and evaluation of the accuracy of the magnetic OMP method
are also performed in Sect. \ref{Sect:5}, along with a comparison with the conventional center-of-gravity method.
A summary is presented in Sect. \ref{Sect:6}.

\section{Magnetic field estimation}
\label{Sect:2}
Our starting point will be the so-called center-of-gravity (COG) method \citep{Semel67,Rees79}.
Before we describe the COG method, we briefly introduce 
the concept of the weak-field approximation from which we later derive the effective and
apparent longitudinal magnetic field.

\subsection{Disk-Integrated weak-field approximation}
\label{Sect:2.1}
The \emph{local} weak-field-approximation (WFA) \citep[e.g.,][]{Landi92,Stenflo94} 
makes the assumption that the atmosphere of a resolved region 
is permeated by a homogeneous, weak, and depth-independent magnetic field.
Then, a proportionality exists between the observed local Stokes~$V$ profile and 
the spectral derivative of the observed local Stokes~$I$ profile, which can 
be derived from the polarized transfer equation using a first-order Taylor 
expansion of the line profile function \citep[e.g.,][]{Jefferies89}. 
This proportionality or linear relation can be written as
\begin{equation}
V(\lambda) = -g \lambda_B \cos\gamma \frac{dI_{0}(\lambda)}{d\lambda} \: ,
\label{Eq:2.1.1}
\end{equation} 
where $\gamma$ is the angle between magnetic field vector and the line-of-sight
(LOS), $g$ the effective Land\'e factor,
and $\lambda_B$  the Zeeman splitting expressed in wavelength units is given by
\begin{equation}
\lambda_B \: = \: 4.67 \times 10^{-13} g_{\rm eff}\: \lambda_0^2 \: B \: .
\label{Eq:2.1.1b}
\end{equation}
For the sake of brevity, we denote by $\lambda$ 
the relative wavelength shift from the line center $\lambda_0$ hereafter 
and assume that the Stokes profiles are normalized by the local continuum
intensity. 
We moreover introduce the variable $\mu =  \cos\gamma$ and $\alpha = - 4.67 \times 10^{-13} g_{\rm eff} \lambda_0^2 $ to write
Eq. (\ref{Eq:2.1.1}) as
\begin{equation}
V(\lambda) = \alpha B \mu \frac{dI_{0}(\lambda)}{d\lambda} \: = \: \alpha B_l \frac{dI_{0}(\lambda)}{d\lambda}\: ,
\label{Eq:2.1.2}
\end{equation} 
where $B_l$ denotes the longitudinal component of the magnetic field vector.
The proportionality between the observed local Stokes~$V$ and the 
observed local Stokes~$I$ profile
depends on the wavelength. In the local case, i.e. resolved (solar) observations, 
this dependency is usually 
neglected since the same scaling factor, i.e. $\alpha B \mu$, applies for all wavelengths. 
This condition is not generally true for unresolved stellar observations.

Given that the assumptions for the weak-field approximation are applicable,
it is readily seen that for the local and the resolved case, one can obtain an 
estimate for the longitudinal component $B_l$
of the magnetic field through
\begin{equation}
B_l = \frac{V(\lambda)}{\alpha \; dI_{0}(\lambda) / d\lambda} \; .
\label{Eq:2.1.3}
\end{equation} 
For each spectral line, a good estimate up to which magnetic field strength the
WFA remains a valid approximation, can be deduced from the so-called Zeeman saturation \citep{Stenflo94}.
This is the regime from which on the Stokes~$V$ amplitudes no longer grow linearly 
but instead the individual sigma components begin to shift apart according to Eq. \ref{Eq:2.1.1b}. 
The Zeeman saturation regime is reached for the majority of Zeeman sensitive 
spectral lines in the optical at around one kilo-Gauss.

In stellar spectropolarimetric observations, the only observables are the
disk-integrated Stokes profiles.
The obvious requirement for applying the weak-field approximation to stellar spectra is the existence of
a wavelength independent linear scaling between the disk-integrated
Stokes~$V$ profile, denoted hereafter as $\tilde{V}$, and the spectral derivative 
of the disk-integrated Stokes~$\tilde{I}$.
For the following disk integration, we choose a Cartesian coordinate system where the z-axis points along the LOS,
the y-axis is in the plane defined by the LOS and the rotation axis, and the x-axis is perpendicular to that plane. 
Using  $\Delta \lambda_{Rot} = (\lambda_0 \textrm{\textsl{v}} \sin i) / c$, the \emph{disk-integrated} 
weak-field approximation for a rotating star
can then be written in its general form as
\begin{eqnarray}
\tilde{V}(\lambda)= \int\limits_{-1}^{+1} \int\limits_{-\sqrt{1-x^2}}^{+\sqrt{1-x^2}} 
V(B;\lambda-\Delta\lambda_{Rot}x;x,y) \: dx \: dy \: =
\hspace{1.3cm} \nonumber \\
 = \int\limits_{-1}^{+1} \int\limits_{-\sqrt{1-x^2}}^{+\sqrt{1-x^2}} \alpha  B_l(x,y) \: \frac{dI_{0}(\lambda-\Delta \lambda_{Rot}x;x,y)}{d\lambda} \: dx \: dy ,
\label{Eq:2.1.4}
\end{eqnarray}
where we introduced the spatial-dependent longitudinal magnetic field strength $B_l(x,y) = B(x,y) \; \mu(x,y)$.

The observed Stokes $\tilde{V}$ vector depends on the magnetic field, whereas the local 
intensity has no explicit dependency on the magnetic field under the weak-field assumption. The intensity profile 
originates over the entire observable disk. 
This disconnection, in general, prevents applying the WFA to unresolved stellar observation because the disk-integrated profile 
shape of the spectral derivative of Stokes $\tilde{I}$ does not coincide with the observed 
Stokes $\tilde{V}$ profile. However, the assumption that the thermodynamic
parameters in the atmosphere and across the observable disk do not change between magnetic and field-free regions, 
as well as the assumption that the longitudinal magnetic field on the stellar surface has no explicit spatial dependence, 
makes the WFA applicable also to disk-integrated spectra.
In this case one can obtain a similar relation to the one in Eq. (\ref{Eq:2.1.3}) by defining a \emph{mean} longitudinal field strength 
$\langle B_l \rangle,$ which allows one to separate the magnetic field from the integral in Eq. (\ref{Eq:2.1.4}).
The disk-integrated WFA can then compactly written as 
\begin{equation}
\tilde{V}(\lambda) \: = \: \alpha \: \langle B_l \rangle \: \frac{d\tilde{I}(\lambda)}{d\lambda}\: . 
\label{Eq:2.1.5}
\end{equation}

This linear scaling relation must hold for all wavelengths $\lambda$. 
If this is not the case, owing to a spatial dependence of the
underlying magnetic field and rapid rotation, 
the WFA can no longer be applied. The applicability of the disk-integrated WFA is 
also limited to fields below the Zeeman saturation regime. 
Because the disk integration is a linear operation, 
the Zeeman saturation regime, in general, is determined 
not by the \emph{mean} longitudinal field  but 
by the \emph{local} field strengths on the surface of the star. 

\subsection{The effective longitudinal magnetic field}
\label{Sect:2.2}
The center-of-gravity method does in fact bypass the limitation that the observable Stokes~$\tilde{V}$ 
profile must be proportional to the spectral derivative of the Stokes~$\tilde{I}$ profile by using the
first-order moment of the Stokes~$\tilde{V}$ profile \citep{Mathys89}. 
This first-order moment of the Stokes~$V$ signal can be obtained by using Eq. (\ref{Eq:2.1.4}), 
\begin{eqnarray}
\int\limits_{\Lambda} \tilde{V}(\lambda) \: \lambda  \: d\lambda  \: = \hspace{6.0cm} \nonumber \\
= \int\limits_{\Lambda} \int\limits_{-1}^{+1} \int\limits_{-\sqrt{1-x^2}}^{+\sqrt{1-x^2}} \alpha  B_l(x,y) \: 
\frac{dI_{0}(\lambda-\Delta \lambda_{Rot}x;x,y)}{d\lambda} \: dx \: dy \:
\lambda \: d\lambda \: ,
\label{Eq:2.2.1}
\end{eqnarray}
where $\Lambda$ denotes the integration limits that cover the entire
line profile. Instead of deriving the COG method from the weak-line limit we use 
the WFA as a starting point \citep{Semel67}.
Under the WFA, the profile of a spectral line does not alter its 
shape, and we may introduce a line-dependent intensity distribution 
function $\eta$, which accounts for geometrically induced radiative transfer 
effects (e.g., limb darkening and other possible temperature effects).
We then write, for the Stokes~$I$ profile at position $x,y$ on the stellar disk,
\begin{equation}
I(\lambda;x,y) \; = \; \eta(x,y) \; I_0(\lambda;\mu = 1) \; .
\label{Eq:2.2.2}
\end{equation}
Substituting this into Eq. (\ref{Eq:2.2.1}) gives
\begin{eqnarray}
\int\limits_{\Lambda} \tilde{V}(\lambda) \: \lambda \: d\lambda  \: = \hspace{6.0cm} \nonumber \\
= \int\limits_{-1}^{+1} \int\limits_{-\sqrt{1-x^2}}^{+\sqrt{1-x^2}} \alpha  \: B_l(x,y) \: \eta(x,y) \: \frac{dI_{0}(\lambda-\Delta
 \lambda_{Rot}x)}{d\lambda} \: dx \: dy \: \lambda \: d\lambda \: .
\label{Eq:2.2.3}
\end{eqnarray}
 
By taking advantage of the Cartesian coordinate system, where values along the x-axis experience different degrees 
of Doppler shifts $\Delta \lambda$, we can perform the following change of variables, 
$x = \Delta \lambda / \Delta \lambda_{Rot}$, to write
\begin{eqnarray}
\int\limits_{\Lambda} \tilde{V}(\lambda) \: \lambda  \: d\lambda \: = \: \hspace{6.0cm} \nonumber \\
= \: \alpha \int\limits_{\Lambda} \int\limits_{-\Delta\lambda_{Rot}}^{+\Delta\lambda_{Rot}} 
 \int\limits_{-\sqrt{1-(\Delta\lambda/ \Delta\lambda_{Rot})^2}}^{+\sqrt{1-(\Delta\lambda/ \Delta\lambda_{Rot})^2}} 
 B_l(\Delta \lambda / \Delta \lambda_{Rot},y)  \times \hspace{1.0cm} \nonumber \\
 \times \: \eta(\Delta \lambda / \Delta \lambda_{Rot},y) \frac{dI_{0}(\lambda-\Delta\lambda)}{d\lambda} 
 \: dy \: d(\Delta \lambda) \: \lambda\: d\lambda \: .\hspace{1.0cm}
\label{Eq:2.2.4}
\end{eqnarray}
We define the total flux-weighted longitudinal magnetic field distribution $B_l^{\eta}$ along a small
a strip $dx$ or $d(\Delta \lambda)$ of equal velocity as
\begin{eqnarray}
B_l^{\eta}(\Delta \lambda) = 
\int\limits_{-\sqrt{1-(\Delta\lambda/ \Delta\lambda_{Rot})^2}}^{+\sqrt{1-(\Delta\lambda/ \Delta\lambda_{Rot})^2}} 
B_l(\Delta \lambda / \Delta \lambda_{Rot},y) \: \eta(\Delta \lambda / \Delta \lambda_{Rot},y) \: dy .
\label{Eq:2.2.5}
\end{eqnarray}
This allows us to write Eq. (\ref{Eq:2.2.4}) as a convolution integral
\begin{eqnarray}
\int\limits_{\Lambda} \tilde{V}(\lambda) \: \lambda  \: d\lambda  \: = \hspace{5.5cm} \nonumber \\
= \alpha  \int_{\Lambda} 
\int\limits_{-\Delta\lambda_{Rot}}^{+\Delta\lambda_{Rot}} 
 B_l^{\eta}(\Delta \lambda) \: \frac{dI_{0}(\lambda-\Delta\lambda)}{d\lambda} \: d(\Delta \lambda) \: \lambda \: d\lambda \: ,
\label{Eq:2.2.6}
\end{eqnarray}
or in a more compact way 
\begin{eqnarray}
\int\limits_{\Lambda} \tilde{V}(\lambda) \: \lambda  \: d\lambda  
\: = \: \alpha \int\limits_{\Lambda} \frac{d}{d\lambda} \left ( B_l^{\eta} \ast I_{0} \right ) \: \lambda \: d\lambda \: .
\label{Eq:2.2.7}
\end{eqnarray}
The disk-integrated Stokes~$\tilde{I}$ profile is also given by the 
convolution
\begin{equation}
\tilde{I} \: = \: \int\limits_{-\Delta\lambda_{Rot}}^{+\Delta\lambda_{Rot}} \eta(\Delta \lambda) \: I_0(\lambda-\Delta\lambda) \: d(\Delta \lambda) \: ,
\label{Eq:2.2.8}
\end{equation}
where $\eta(\Delta \lambda)$ is defined by
\begin{eqnarray}
\eta(\Delta \lambda) = 
\int\limits_{-\sqrt{1-(\Delta\lambda/ \Delta\lambda_{Rot})^2}}^{+\sqrt{1-(\Delta\lambda/ \Delta\lambda_{Rot})^2}} 
\eta(\Delta \lambda / \Delta \lambda_{Rot},y) \: dy .
\label{Eq:2.2.9}
\end{eqnarray}

With this definition we may separate a mean longitudinal magnetic field from Eq. (\ref{Eq:2.2.5}) 
that we call the effective longitudinal magnetic field $B_{\rm eff}$. 
Using Eqs. (\ref{Eq:2.2.8}) and (\ref{Eq:2.2.9}), we can rearrange Eq. (\ref{Eq:2.2.6}) to obtain
the following expression for the effective longitudinal magnetic field 
\begin{equation}
B_{\rm eff} = \frac{\int_{\Lambda} \tilde{V}(\lambda) \: \lambda \:d\lambda} {\alpha \int_{\Lambda} 
\left ( d\tilde{I}(\lambda) / d\lambda \right ) \: \lambda \: d\lambda} \; .
\label{Eq:2.2.10}
\end{equation}
Performing the integration by parts in the denominator and keeping in mind that
we implicitly deal with normalized profiles, we obtain
\begin{equation}
B_{\rm eff} = \frac{\int_{\Lambda} \tilde{V}(\lambda) \: \lambda \:d\lambda} {\alpha  W_I} \; ,
\label{Eq:2.2.11}
\end{equation}
where $W_I$ is the equivalent width of the disk-integrated Stokes $I$ profile.
This is the COG method for disk-integrated observations 
introduced for solar observations by \citet{Semel67,Rees79}. 

Making a transformation of the \emph{relative} wavelength according to 
$\lambda \rightarrow \textrm{\textsl{v}} \lambda_0 / c$, 
we obtain the following equation for the velocity domain
\begin{equation}
B_{\rm eff} = \frac{\lambda_0 \: \int_{V} \tilde{V}(\textrm{\textsl{v}}) \: \textrm{\textsl{v}} \: d\textrm{\textsl{v}}} {c \: \alpha  \: W_I} 
= -2.14 \times 10^{12} \: \frac{\int_{V} \tilde{V}(\textrm{\textsl{v}}) \: \textrm{\textsl{v}} \: d\textrm{\textsl{v}}} {c \lambda_0 g \: W_I} \; ,
\label{Eq:2.2.12}
\end{equation}
where the central wavelength $\lambda_0$ is given in $\AA$.
This form is extensively used in spectropolarimetric observations where the 
spectral line profiles are often preprocessed
by transforming them into the velocity domain during a reconstruction 
process \citep[see e.g., ][]{Donati97,Wade00}.

\subsection{The apparent longitudinal magnetic field}
\label{Sect:2.3}
If we now consider the case where the projected rotational velocity of the star is sufficiently large and the effective 
longitudinal magnetic field is zero owing to a balance of positive and negative 
polarities, the disk-integrated Stokes~$V$ signal of the star does not 
necessarily cancel out and vanishes. In fact, the net absolute circular 
polarization; i.e., the integral of the absolute value of the observed 
Stokes~$V$ signal over wavelength is in general 
not zero. Even though we measure no effective field,
the polarized profile is clearly detectable and \emph{apparently} a magnetic field must be present. 
Of course, this is one of the effects that is utilized by ZDI
to resolve surface magnetic fields on rapidly rotating stars.
Having such a nonvanishing net absolute circular polarization means that the 
magnetic field exhibits a flux imbalance for 
each or some of the resolved surface areas on the stellar disk,
even though the overall disk-averaged magnetic field is 
flux balanced.

To characterize this imbalance, we define the apparent longitudinal magnetic field.
We start by defining the integrated Stokes~$V^*$ along a small iso-radial velocity 
strip $d(\Delta \lambda)$ as
\begin{equation}
V^*(\Delta \lambda) \: = \: 
\int\limits_{-\sqrt{1-(\Delta\lambda/ \Delta\lambda_{Rot})^2}}^{+\sqrt{1-(\Delta\lambda/ \Delta\lambda_{Rot})^2}} 
V(\Delta \lambda / \Delta \lambda_{Rot},y) \:  dy \: .
\label{Eq:2.3.1}
\end{equation}
The disk-integrated first-order moment of the Stokes~$V$ profile can then be written as
\begin{equation}
\int\limits_{\Lambda} \tilde{V}(\lambda) \: \lambda  \: d\lambda   
\: = \: \int\limits_{\Lambda} \int\limits_{-\Delta\lambda_{Rot}}^{+\Delta\lambda_{Rot}} V^*(\Delta \lambda) \: d(\Delta \lambda) \: \lambda \: d\lambda \: .
\label{Eq:2.3.2}
\end{equation}
The total amount of the effective longitudinal magnetic field can be expressed by
\begin{equation}
B^{tot}_{\rm eff} \: = \: \frac{1}{\alpha  \: W_I} \: \int\limits_{-\Delta\lambda_{Rot}}^{+\Delta\lambda_{Rot}} 
\left | \int\limits_{\Lambda}  V^*(\Delta \lambda) \: \lambda \: 
d\lambda \right | \: d(\Delta \lambda)\: .
\label{Eq:2.3.3}
\end{equation}
This is, of course, not an observable, but assuming that we have a number of $n$ 
discrete and spectroscopically resolvable
radial-velocity strips across the surface of the star that produce separated 
Stokes~$V$ profiles, we may write
\begin{eqnarray}
\frac{1}{\alpha  \: W_I} \int\limits_{-\Delta\lambda_{Rot}}^{+\Delta\lambda_{Rot}}\left | \int\limits_{\Lambda}  V^*(\Delta \lambda) \: \lambda \: d\lambda \right | \: d(\Delta \lambda)\:
\: \geq \: \hspace{2.5cm} \nonumber \\ 
\geq \: \frac{1}{\alpha  \: W_I} \sum\limits_{i=0}^{n} \left | \int\limits_{\Lambda} \hat{V}^*_i(\lambda) \: \lambda \: d\lambda \right | \: ,
\label{Eq:2.3.4}
\end{eqnarray} 
where $\hat{V}^*_i$ is the Stokes~$V$ profile originating in the $i$-th 
resolved radial-velocity strip. 
In this case the observed Stokes~$V$ profile is the composition of $n$ individual
contributions coming from different velocity-resolved regions. 
Using Eqs. (\ref{Eq:2.3.4}) and (\ref{Eq:2.3.2}), we can finally define the apparent
longitudinal magnetic field as 
\begin{equation}
B_{\rm app} = \frac{\sum\limits_{i=0}^{n} \left | \int\limits_{\Lambda} \hat{V}^*_i(\lambda) \: \lambda \: d\lambda \right |} {\alpha  W_I} \; ,
\label{Eq:2.3.5}
\end{equation}
or in velocity coordinates
\begin{equation}
B_{\rm app} = \frac{\lambda_0 \: \sum\limits_{i=0}^{n} \left |  \int_{V} \tilde{V}_i(\textrm{\textsl{v}}) \: \textrm{\textsl{v}} \: d\textrm{\textsl{v}} \right |} 
{c \: \alpha  \: W_I} \: .
\label{Eq:2.3.6}
\end{equation}
The apparent longitudinal magnetic field measures the maximum resolved field that can be
obtained from the observation. In general, we have $B^{tot}_{\rm eff} \ge B_{\rm app}$, and the amount of 
the $B_{\rm app}$ depends on the intrinsic line width of the local Stokes~$I$ 
profile. However, it can
give an estimate for the distribution of the absolute effective field over the stellar disk. 
Although it is not clear in the first place how to obtain the apparent longitudinal field and
the flux density distribution from the
disk-integrated Stokes~$V$ profile, we present an easy and fast way 
for estimating this quantity in the next section. 

\section{Sparse approximation of Stokes profiles and magnetic field detection}
\label{Sect:3}
The basic idea of our proposed approach is to find an accurate approximation 
of the observed Stokes~$V$ profile by decomposing the original signal with a
set of well chosen elementary waveforms. The stagewise approximation by
these elementary basis functions
allows us to apply the equations for the effective and apparent longitudinal magnetic field
to each of these elementary functions and eventually obtain the desired magnetic quantities.
The decomposition we describe in the following is based on a prescribed overcomplete set 
of elementary functions, called signal or profile atoms.
Overcomplete here means that the number of elements provided by the dictionary is 
redundant, i.e. larger than the actual dimension of the signal space.
In contrast to many nonredundant (e.g., orthogonal) transformations, a linear expansion 
of a given signal profile with an overcomplete set of signal atoms 
often facilitates a more efficient and sparse approximation \citep{Tropp04,Donoho06}.
By using a suitable and redundant set of profile atoms, 
we show in the following that a sparse approximation of 
observed Stokes line profiles, allows a resolved analysis of the 
effective and apparent longitudinal magnetic flux density.

\subsection{Orthogonal matching pursuit}
\label{Sect:3.1}
The matching pursuit (MP) algorithm is an iterative algorithm for adaptive signal reconstruction and approximation.
A given signal or line profile is decomposed into a linear expansion of dictionary elements. 
The actual realization of the dictionary is discussed in
Sect. \ref{Sect:3.2}. For the moment it suffices
to consider the dictionary as a collection of possibly linear-dependent
elementary waveforms. 
We begin with a description of the 
original MP algorithm introduced by \citet{Mallat93} and
then of its improvement called OMP \citep{Pati93}, which we
use for our analysis.

Following \citet{Mallat93} let us define a dictionary $D(\vec{x})$ as a collection of a large number of elementary wave functions 
$[\vec{x}_1,\vec{x}_2,...,\vec{x}_n]$  formally defined in Hilbert space
$\mathcal{H}$ where each vector is of unit norm $\| \vec{x}_i \| = 1$. 
A given signal $f \in \mathcal{H}$ is approximated by MP in a first step by projecting $f$ onto a vector
$x_l \in D$ such that 
\begin{equation}
 f \: = \: \langle f,x_l \rangle x_l \: + \: Rf \: ,
\label{Eq:3.1.1}
\end{equation}
where $Rf$ is the current residual of the approximation. Since the residual $Rf$
of the current estimate is orthogonal to $x_l$, we have the following energy conservation
\begin{equation}
\| f \|^2 \: = \: | \langle f,x_l \rangle |^2 \: + \: \|Rf\|^2 \: . 
\label{Eq:3.1.2}
\end{equation}
To minimize $\|Rf\|^2$ the vector $x_l \in D$ is chosen such that 
the maximum projection (i.e., correlation) is found between the residual vector and
all dictionary atoms,
\begin{equation}
x_l \: = \: \arg\max_{x_k \in D} | \langle R f,x_k \rangle |  \: .
\label{Eq:3.1.3}
\end{equation}
The algorithm now iteratively decomposes the residual $Rf$
by repeated projections onto the dictionary atoms. The recursive algorithm can be written in a compact way
if we set the initial values for the residual $R^0f = f$ and the initial approximation to $f_0 = 0$,
\begin{eqnarray}
(I) \: \: \: x_{l_n} \: = \: \arg\max_{x_k \in D} | \langle R^nf,x_k \rangle | \: , \hspace{3.5cm} \nonumber \\
(II) \: \: f_{n+1} \: = \: f_n \: + \: \langle R^nf,x_{l_n} \rangle x_{l_n} \:  , \hspace{3.3cm}\nonumber \\
(III) \: \: R^{n+1}f \: = \: R^nf - \: \langle R^nf,x_{l_n} \rangle x_{l_n} \: .\hspace{2.8cm}
\label{Eq:3.1.4}
\end{eqnarray}
An increasingly closer approximation to the signal is obtained by repeating steps (I) to (III).
A proof of convergence for the MP algorithm can be found in \citet{Mallat93}.

The MP algorithm can be improved by realizing that the steps taken by the MP are 
not optimal in the sense that a newly chosen dictionary vector is not orthogonal 
to the previously selected vectors \citep{Pati93}. This can be resolved by orthogonalizing the newly selected vector
relative to the $n-1$ previously selected vectors. The orthogonalization, which is essentially a Gram-Schmidt procedure
\citep{Bronstein97}, uses an auxiliary vector $z$ that expresses the yet unexplained component  
of the newly selected vector $x_{l}$. For the $n$-th iteration we may write
\begin{equation}
z_{n} \: = \: x_{l_n} \: - \sum_{p=0}^{n-1} \frac{\langle x_{l_n},z_{p} \rangle z_{p}}{\| z_{p} \|^2} \: ,
\label{Eq:3.1.5}
\end{equation}
where the initial condition is set to $z_{0} = x_{l_0}$. 
The new residual $R^{n+1}f$ can then be expressed by the projection of the current residual $R^{n}f$ on $z_{n}$,
\begin{equation}
R^{n+1}f \: = \: R^nf - \: \frac{\langle  R^nf,z_{n} \rangle z_{n}}{\| z_{n} \|^2} \: ,
\label{Eq:3.1.6}
\end{equation}
using the relation in Eq. (\ref{Eq:3.1.5}). The latter can also be written as
\begin{equation}
R^{n+1}f \: = \: R^nf - \: \frac{\langle  R^nf,x_{l_n} \rangle z_{n}}{\| z_{n} \|^2} \: .
\label{Eq:3.1.7}
\end{equation}
Compared to the ordinary MP algorithm a component is now subtracted in a direction that is
orthogonal to all previously selected vectors.
Beginning with the initial conditions $R^0f = f$ and $f_0 = 0$ as well as $z_{0} = x_{l_0}$  
the recursive algorithm for the OMP can then be expressed as
\begin{eqnarray}
(Ia) \: \:x_{l_n} \: = \: \arg\max_{x_k \in D} | \langle R^nf,x_k \rangle | \: , \hspace{3.8cm} \nonumber \\
(Ib) \: z_{n} \: = \: x_{l_n} \: - \sum_{p=0}^{n-1} \frac{\langle x_{l_n},z_{p} \rangle z_{p}}{\| z_{p} \|^2} \: ,\hspace{3.7cm} \nonumber \\
(II) \: \:f_{n+1} \: = \: f_n \: + \: \frac{\langle  R^nf,x_{l_n} \rangle z_{n}}{\| z_{n} \|^2} \: ,\hspace{3.7cm} \nonumber \\
(III) \: \: R^{n+1}f \: = \: R^nf - \: \frac{\langle  R^nf,x_{l_n} \rangle z_{n}}{\| z_{n} \|^2} \: . \hspace{3.2cm}
\label{Eq:3.1.8}
\end{eqnarray}
After $n$ iterations we obtain the following sparse approximation of our signal,
\begin{equation}
f \: \approx \: \sum_{n} \frac{\langle  R^nf,x_{l_n} \rangle \: z_{n}}{\| z_{n} \|^2} \: .
\label{Eq:3.1.8}
\end{equation} 
A convergence analysis and more results for the OMP algorithm 
can be found in \citet{Davis97}.
The set of selected signal atoms $\{ x_{l_0},x_{l_1},...,x_{l_n} \}$, its corresponding projection weights, 
and auxiliary vectors give us the opportunity to analyze any given signal in terms of a small number 
of dictionary basis functions.

\subsection{The wavelet dictionary for Stokes profile approximation}
\label{Sect:3.2}  
To facilitate a sparse and compact representation of a given Stokes profile, 
we need a set of basis functions that 
closely resemble the individual building blocks of a disk-integrated circular polarized 
profile. These building blocks are the local Stokes~$V$ profiles originating in a 
small resolved surface area of the star.
From the definition of a general spectral line profile \citep{Gray05}, we know 
that a Gaussian function can provide a reasonably good approximation of a spectral absorption profile
in cases where damping is not too strong.
Under the regime of the WFA, where the Stokes~$V$ profile is proportional to
the derivative of the intensity profile, it seems obvious that an appropriate 
and easy way to approximate a composite (i.e., disk-integrated) 
Stokes~$V$ profile can be achieved by using a dictionary of the derivatives of Gaussian functions. 
In fact, we use a wavelet frame 
constructed from scaled and translated versions of first derivatives of a 
Gaussian \citep{Torrence98}. 
We define the dictionary $D$
as a set of elementary functions or atoms $\Psi_{j,k}$ of unit norm as
\begin{equation}
D \: = \: \{ \Psi_{j,k}, i,k \in Z \} \: . 
\label{Eq:3.2.1}
\end{equation}
The atoms are constructed by a set of scaled and translated versions of a mother wavelet $\Psi$.
For the mother wavelet we choose, as mentioned above, the first derivative of a Gaussian function,
\begin{equation}
\Psi(x) \: = \: - \: \left ( \frac{2}{\sqrt{\pi}} \right )^{1/2} \: x \: e^{x^2/2} \: .
\label{Eq:3.2.2}
\end{equation}
The wavelet function obtained by a scaled and translated version of the mother wavelet can be 
expressed in the wavelength domain as
\begin{equation}
\Psi_{j,k}(\lambda) \: = \: s_{j}^{-\frac{1}{2}} \: \Psi \left ( \frac{\lambda-\lambda_k}{s_j} \right ) \: ,
\label{Eq:3.2.3}
\end{equation}
where $s_j$ and $\lambda_k$ denote the scaling and translation parameter.
To obtain a discrete set of wavelet functions, we use the following discretization of the scale parameter
\begin{equation}
s_j \: = \: s_0 2^{j \Delta_r } \: , \mbox{where} \: j=0,1,...,L \: ,
\label{Eq:3.2.4}
\end{equation} 
where $s_0$ defines the smallest resolvable scale, which we set to $2\delta \lambda$, 
twice the wavelength step of the observations or synthetic calculations.
The parameter $\Delta_r$ is a variable value that provides the resolution of the transform 
and which is set throughout this paper to $\Delta_r = 0.125$. This
provides a reasonable trade-off between resolution and computational cost \citep[e.g.,][]{Torrence98}. 
The largest scale $L$ is determined according to
\begin{equation}
L \: = \: \frac{ log_2(N \: \delta \lambda / j_0) } { \Delta_r } \: ,
\label{Eq:3.2.5}
\end{equation}
where $N$ is the number of wavelength points in the line profile.
With the wavelet function Eq. (\ref{Eq:3.2.3}), we can express the wavelet coefficients of an 
observed spectrum $V(\lambda)$ as the dot product between the Stokes $V$ profile and the 
wavelet function, 
\begin{equation}
w_{j,k} \: = \: \int_R V(\lambda) \: \Psi_{j,k}(\lambda) \: d\lambda \: = \: \langle V,\Psi_{j,k} \rangle \: ,
\label{Eq:3.2.5}
\end{equation}
where the integration limits $R$ are large enough to cover the entire profile.
The projections (i.e., dot products) of Eq. (\ref{Eq:3.1.8}) used in the OMP algorithm 
can be computed easily by calculating the convolution between the current residual and the wavelet function 
for all scales $s_j$. 

The decomposition by a restricted set of wavelets with different scales 
provides a multiscale representation of our original signal profile. 
We use this multiscale decomposition to apply an approach called multiresolution support \citep{Starck06} to complement 
the OMP algorithm with an efficient detection procedure. 

\subsection{Magnetic field detection}
\label{Sect:3.3}
We first start with the assumption that 
our observation vector (i.e., spectral line profile) $V$ contains the true 
signal vector $S$ and additive noise $N$ such that we may write, 
for the observed signal,
\begin{equation}
V(\lambda) \:  = \: S(\lambda) \: + \:  N(\lambda) \: .
\label{Eq:3.3.1}
\end{equation}
Separating the signal from noise can be cast in the framework of nonparametric signal estimation. One very
successful way of reconstructing an unknown signal from noisy observation is thresholding introduced and theoretically
investigated, e.g., by \citet{Donoho95}. For our detection analysis we take a similar thresholding approach that
uses the multiscale decomposition of the signal to test the significance of every wavelet coefficient individually.
The basic strategy of shrinkage or thresholding methods is to consider the observed signal in 
a transformed (e.g., wavelet) domain rather than in the original data domain.
Similar to Fourier filtering, wavelet-based detections methods rest on the idea that the noise 
contribution exhibits a different statistical behavior in the transformed domain.

The OMP algorithm approximates the observed line profile in an iterative process by projecting the dictionary atoms 
onto the current residual of the observation while picking the 
most correlated signal atom in each iterative cycle.
Using a wavelet dictionary, the projections can be efficiently 
calculated by a convolution, i.e., wavelet transform of the observed line profile.
These projections between an individual wavelet function of scale $j$ at position $k$ and the current residual 
of the observation $V$ in the presence of noise can be
written thanks to the linearity of the wavelet transform as
\begin{equation}
\langle V,\Psi_{j,k} \rangle \:  = \: \langle S,\Psi_{j,k} \rangle \: + \:  \langle N,\Psi_{j,k} \rangle \: ,
\label{Eq:3.3.2}
\end{equation}
which we may write according to Eq. (\ref{Eq:3.2.4}) as the wavelet coefficient $w_{j,k}$
\begin{equation}
w_{j,k}^V \: = w_{j,k}^S \: + \: w_{j,k}^N \: .
\label{Eq:3.3.3}
\end{equation}
When including the position index $k$ into a vector notation, this can be written in a compact way by using an 
operator or matrix notation
\begin{equation}
\vec{W}_V(j) = \vec{W}_S(j) + \vec{W}_N(j) \: ,
\label{Eq:3.3.4}
\end{equation}
where $\vec{W}_V$,$\vec{W}_S$, and $\vec{W}_N$ represent the convolution over the wavelength index $k$ on a specific scale $j$.
The expectation value for the wavelet transform of the noise contribution is $E\{\vec{W}_N\}=0,$ and the covariance
is given by $\Sigma = \vec{W}_N\vec{W}_N^T$, where T denotes the transpose. Uncorrelated white noise 
reduces the covariance matrix $\Sigma$ to a diagonal matrix. Unless the matrix is orthogonal (i.e., by using an 
orthogonal set of basis functions), we obtain for each scale $j$ a different noise level $\sigma_j$.
\begin{figure*}[!t]
\begin{minipage}{\textwidth}
\centering
\includegraphics[width=\textwidth]{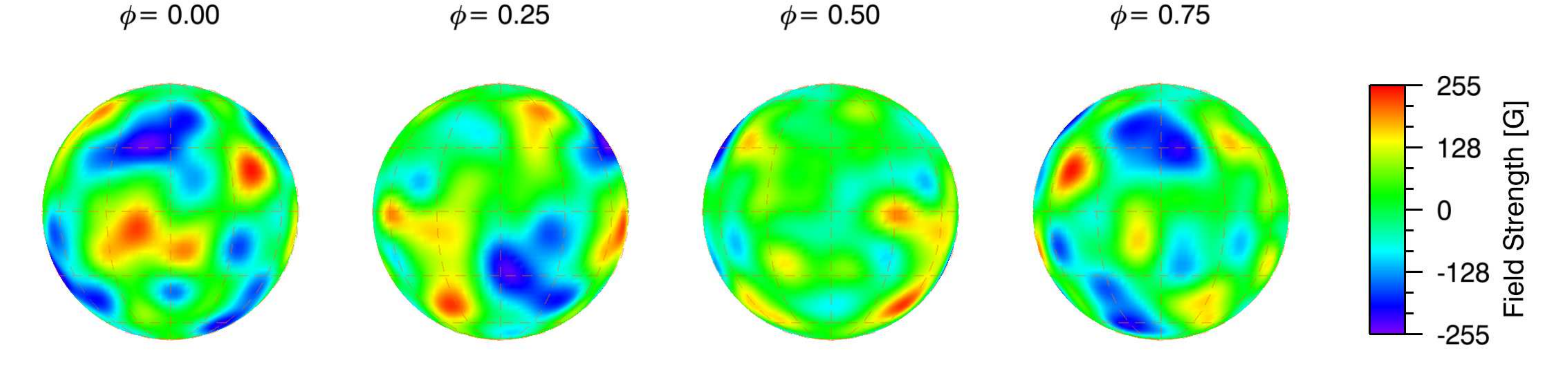}
\caption{Orthographic maps of the random magnetic distribution of
the test star at four different rotational phases $\phi$.}
\label{Fig:1}
\end{minipage}
\end{figure*}
Before we address the problem of estimating the noise level on each resolution scale $j$, we take a closer look at 
the detection problem. The wavelet transform yields a level- or resolution-dependent representation of the observed 
spectral line profile in terms of the wavelet coefficients $w_{j,k}$. A decision
about whether this wavelet coefficient is significant, i.e., whether it includes signal information or not, can be cast in 
a hypothesis-testing framework.
For this, we can state the null hypothesis $H_0$ such that $w_{j,k}$ contains only noise and no signal information. 
Whether the null hypothesis will be rejected or not depends on probability $P_n$,
\begin{equation} 
P_n \:  = \: Prob(\| w_{j,k} \| < \tau | H_0) \: ,
\label{Eq:3.3.5}
\end{equation}
where $\tau$ is a detection threshold. The null hypothesis will be rejected if $P_n$ is smaller than
a given significance level $\epsilon$; i.e. $P_n(\tau) < \epsilon$ . 
For a Gaussian noise distribution with zero mean and standard deviation $\sigma$, we may write the 
probability density for $w_{j,k}$ as 
\begin{equation} 
p(w_{j,k}) \: = \: \frac{1}{\sqrt{2\pi} \sigma} \: e^{\frac{-w_{j,k}^2}{2\sigma^2_j}} \: .
\label{Eq:3.3.6}
\end{equation}    
The probability $P_n$ of rejecting $H_0$ can then be calculated by integrating over all weights $w_{j,k}$ 
that are greater than a threshold $\tau$,
\begin{equation} 
P_n^{+} \: = \: \frac{1}{\sqrt{2\pi} \sigma} \: \int_{\tau}^{+\infty} \: e^{\frac{-w_{j,k}^2}{2\sigma^2_j}} \: dw_{j,k} \: .
\label{Eq:3.3.7}
\end{equation} 
Since the wavelet transform of the Stokes profile can have both positive and negative values, we also 
need to integrate over all negative weights $w_{j,k}$ up to the threshold value $-\tau$
\begin{equation} 
P_n^{-} \: = \: \frac{1}{\sqrt{2\pi} \sigma} \: \int^{-\tau}_{-\infty} \: e^{\frac{-w_{j,k}^2}{2\sigma^2_j}} \: dw_{j,k} \: .
\label{Eq:3.3.8}
\end{equation} 
When assuming stationary noise and choosing a specific significance level $\epsilon$, this reduces to the following 
decision or threshold detection rule \citep[e.g.,][]{Starck06}
\begin{eqnarray} 
\|w_{j,k} \| \: \geq k\sigma_j \: \mbox{then $w_{j,k}$ is significant} \nonumber \hspace{0.7cm} \\ 
\|w_{j,k} \| \: \leq k\sigma_j \: \mbox{then $w_{j,k}$ is not significant} \: ,
\label{Eq:3.3.9}
\end{eqnarray}
where $k$ depends on the value of $\epsilon$. Choosing $\epsilon = 0.0027$ results in $k=3,$ the well known $3\sigma$
detection threshold.   
This thresholding detection scheme can be introduced readily into the OMP algorithm of Sect. \ref{Sect:3.1}
by implementing the threshold detection rule of Eq. (\ref{Eq:3.3.9}) as an additional step (Ib) after calculating the 
maximum projection (i.e. inner product) in step (I). If the currently selected wavelet coefficient
is not significant, it will be deleted from the dictionary for this iteration cycle, and the algorithm restarts the
current iteration cycle with step (I). 
The noise level $\sigma_j$ for each scale $j$ can be obtained by a wavelet transform of simulated noise.
This process yields level-dependent thresholds and has the advantage that 
different types of noise (e.g., Gaussian or Poisson), as well as correlated noise, can be modeled.

\section{Application to synthetic observations}
\label{Sect:4}
We now use the OMP algorithm to estimate the effective and apparent longitudinal magnetic field.
Before we begin with a statistical analysis of the accuracy of our proposed method,
we give an illustrative example to highlight the functionality of the algorithm.
The decomposition of an observed Stokes~$V$ profile into the building blocks
of the wavelet dictionary provides the opportunity for each selected signal or profile atom to 
be individually analyzed in terms of the effective and apparent longitudinal magnetic field. 
Thanks to the linearity of the algorithm, we can evaluate 
Eqs. (\ref{Eq:2.2.11}) and Eq. (\ref{Eq:2.3.5}) for each signal atom and iteratively approximate the 
desired magnetic quantities. Each contributing signal atom found by the OMP algorithm
identifies a resolved structure in the wavelength (or velocity) domain. The overall number of 
iterations then also determines the summation index $n$ for the apparent longitudinal field 
in Eq. (\ref{Eq:2.3.5}).
The stopping criterion is given by the detection threshold as soon as the approximation enters the noise 
and no more significant signal atoms are detected.

In the following example we have built a stellar surface model with our \emph{iMap} code \citep{Carroll12}.
We created a random distribution of a radial magnetic field; i.e.,
each surface segment has a Gaussian random distribution of its radial magnetic field strength with a 
mean value centered at zero Gauss and a standard deviation of 500 Gauss. 
The effective temperature of the test star is 5250 K, and it has 
solar abundance and a gravity of log$(g)=4.0$. The projected rotational velocity $v \sin i$ is $35$ km\,s$^{-1}$,
and micro- and macroturbulence were set to 2 km\,s$^{-1}$. The inclination of the rotational axis 
is 90$^\circ$.
Because the original surface-grid resolution with a 5$^\circ$ by 5$^\circ$ segmentation creates a strong cancellation 
of the Stokes~$V$ signal we decided, for this example case, to additionally smear out the distribution over the 
surface with a Gaussian filter that has an angular spread of 15$^\circ$. 

\begin{figure}
\centering
\includegraphics[width=9cm]{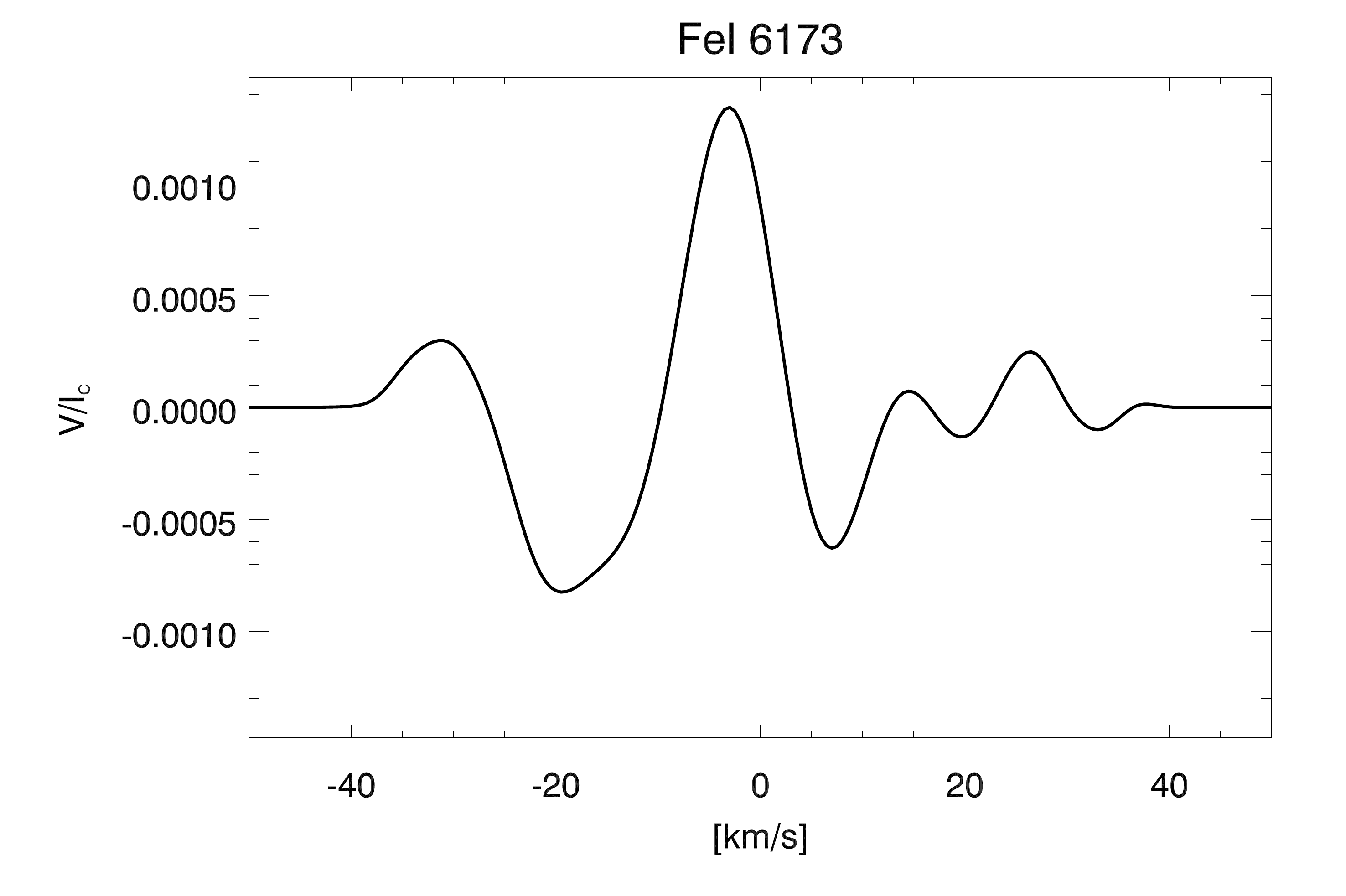}
\caption{Stokes~$V$ of the Zeeman sensitive iron line FeI 6173 for the
random field distribution of the test star at phase 0.0.} 
\label{Fig:2}
\end{figure}

\begin{figure*}[!t]
\begin{minipage}{\textwidth}
\centering
\includegraphics[width=9cm]{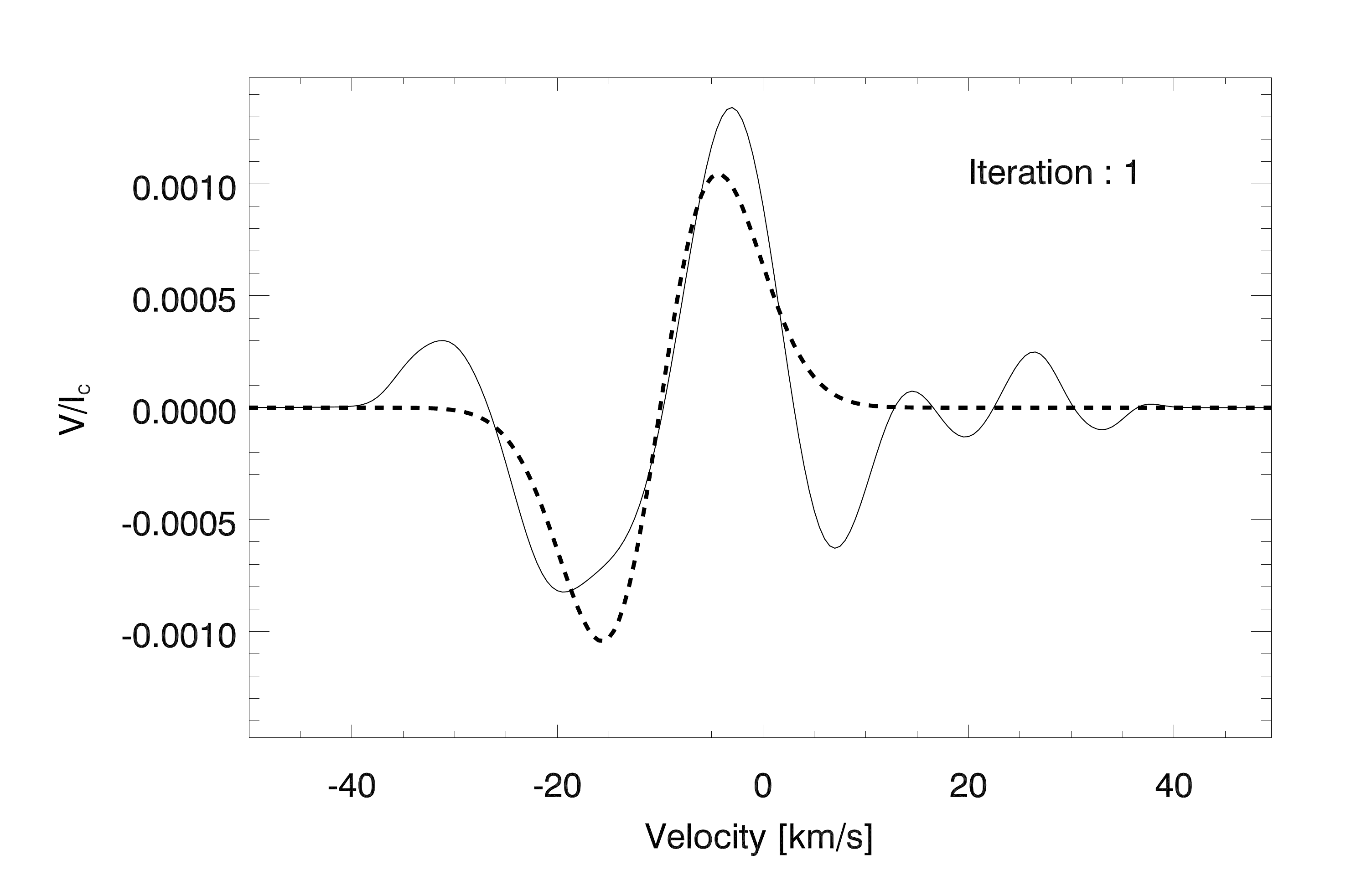}
\includegraphics[width=9cm]{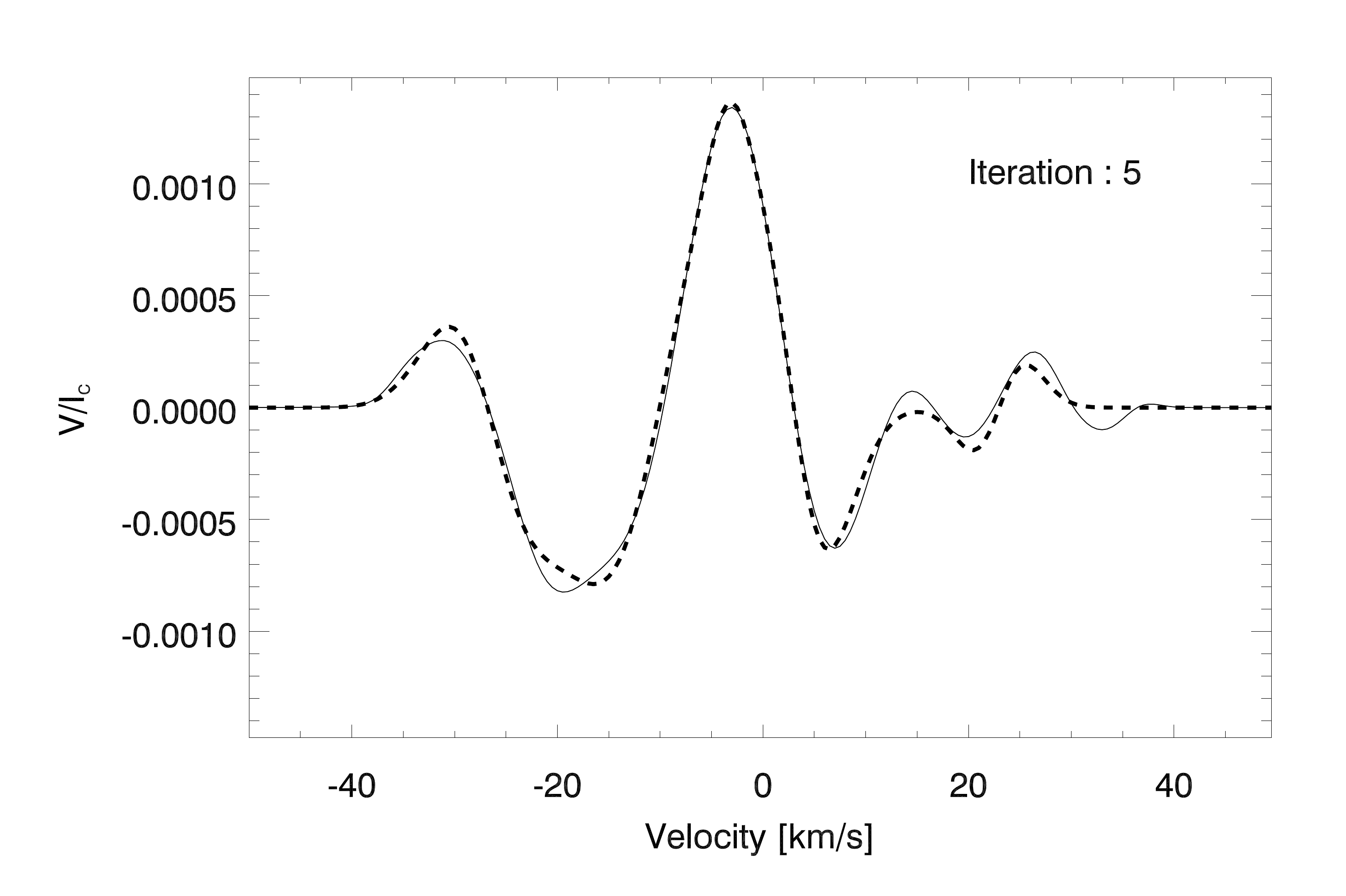}
\includegraphics[width=9cm]{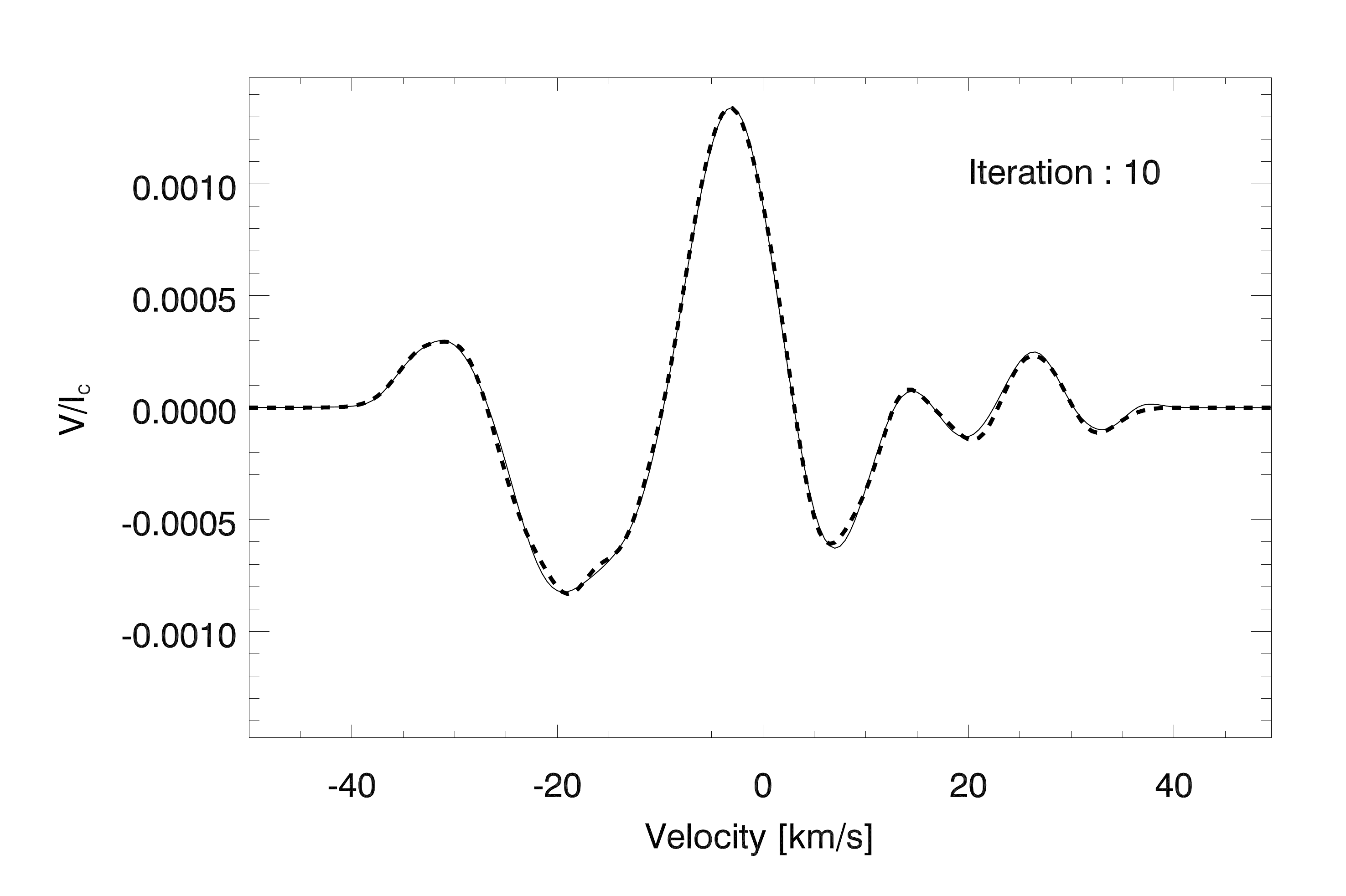}
\includegraphics[width=9cm]{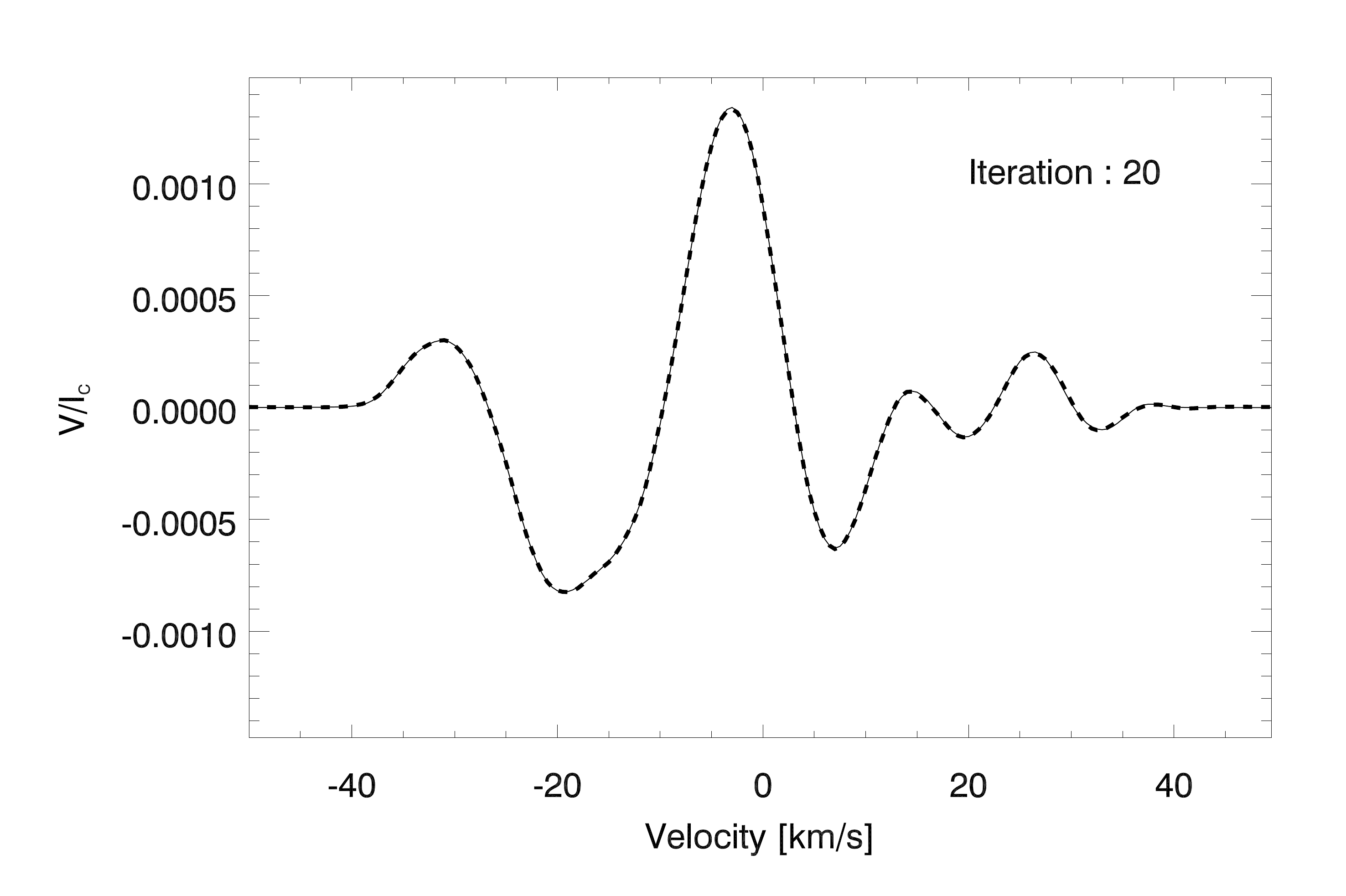}
\end{minipage}
\caption{Magnetic OMP approximation at different iteration cycles. The true synthetic 
profile (solid line) and the approximation of the OMP method (dashed line).}
\label{Fig:3}
\end{figure*}

\begin{figure}
\centering
\includegraphics[width=9cm]{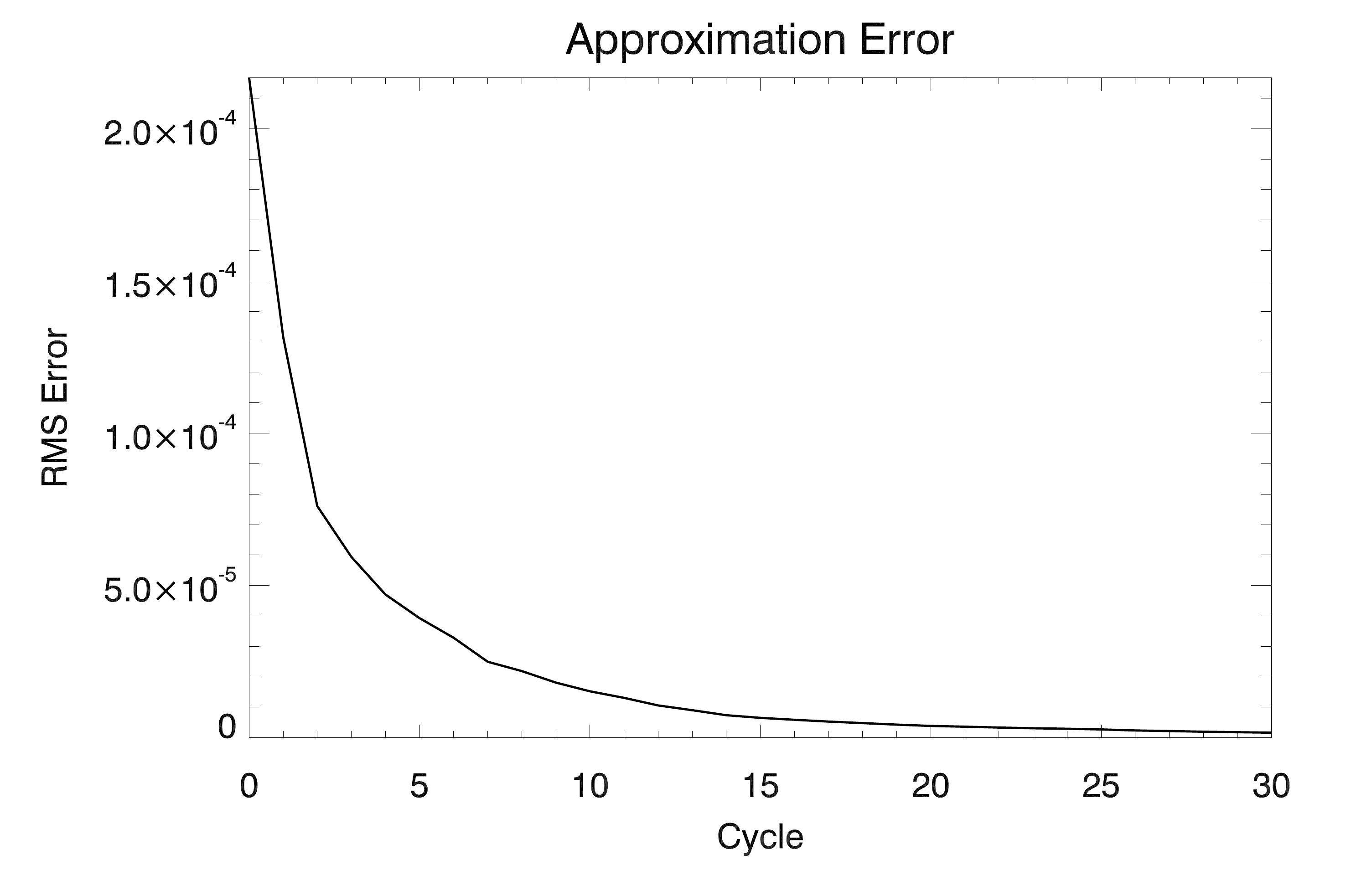}
\caption{The root mean square approximation error over the iteration cycle.} 
\label{Fig:4}
\end{figure}

This smoothing leads to large random clusters of magnetic fields over the surface.
The resulting magnetic field surface distribution is shown in Fig.~\ref{Fig:1}. 
For the synthetic calculation, we used the magnetically sensitive iron line at 6173 $\AA$. 
Line parameters were taken from the VALD line database \citep{Piskunov95,Kupka99}.
The resulting synthetic Stokes~$V$ profile for phase 0.0 is shown in Fig.~\ref{Fig:2}. 
To highlight the performance of the method, we used noise-free Stokes profiles in this
first test. The stopping criterion for the noise-free
case was substituted by a convergence criterion where the iteration was stopped as soon as there 
was no significant improvement in the root-mean-squared (RMS) error of the approximation.
The approximations are illustrated in Fig.~\ref{Fig:3} where four snapshots at different stages
in the approximation process are shown. 
The magnetic OMP algorithm gradually identifies the most coherent signal atoms on different scales
and positions before it finally approximates the entire observed Stokes~$V$ profile 
with a linear combination of all selected dictionary atoms. 
As can be seen in Fig.~\ref{Fig:3},  
the algorithm finds the one best-matching signal atom in the first iteration (upper left). 
In the approximations 5 (upper right), 10 (lower left), and 20 (lower right) 
one can follow the rapid improvement after adding more and more signal atoms to the residuals. 
Already at iteration cycle 10 (i.e., 10 dictionary atoms), the differences between the original 
synthetic observation and the approximation is hardly visible.
In Fig.~\ref{Fig:4} the RMS error of the approximation is plotted over the iteration cycle, 
which also demonstrates the rapid convergence of the magnetic OMP algorithm.
However, the approximation of the pure Stokes~$V$ profile is not the main task of the magnetic OMP method here, 
because it is supposed to give accurate estimates of the effective and apparent magnetic field as well.

\begin{figure}
\centering
\includegraphics[width=9cm]{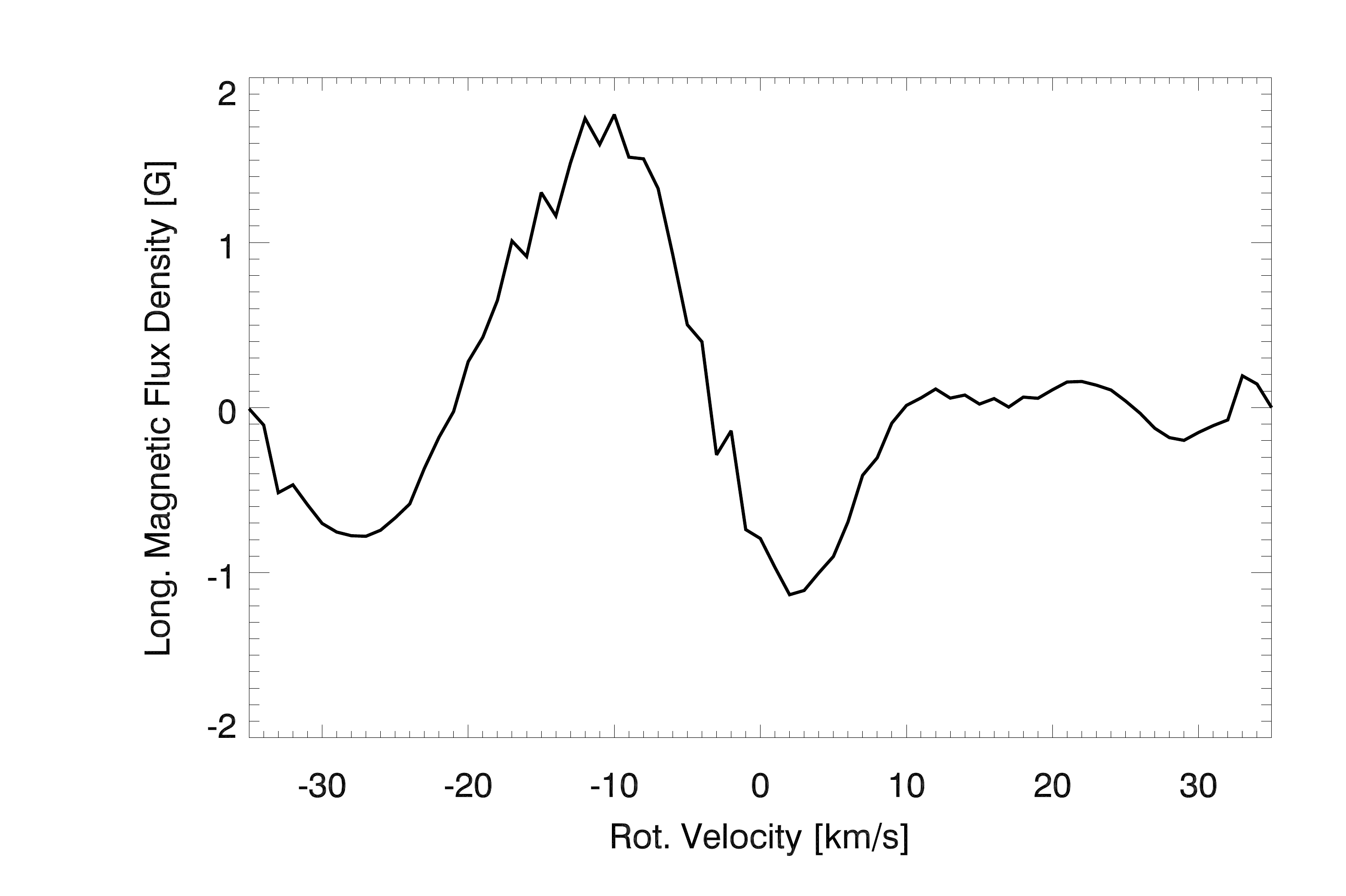}
\caption{Distribution of the longitudinal magnetic flux density across the rotational velocity for phase 0.0 of the
test star.} 
\label{Fig:5}
\end{figure}

The effective and apparent longitudinal magnetic field is estimated during the approximation process 
from each contributing signal atom. 
To compare the estimated values with the \emph{true} values, we need to extract the true effective longitudinal
magnetic field from the model star. This is straightforward and only requires projecting the magnetic field vector 
of each surface segment onto the line-of-sight and eventually to sum the area-weighted 
contribution up from all visible surface segments. The extraction of the true
apparent longitudinal magnetic field value from the model deserves some more explanation. In principle we could 
use the same process as for the effective longitudinal field and sum the absolute value up instead of the plain values
from the stellar model,
but this would not take the cancellation of the Stokes~$V$ signals coming from within iso-radial velocity strips into account. 
We would therefore overestimate the apparent field; in fact, we would retrieve the total absolute 
effective longitudinal magnetic field. 
To obtain an adequate value for the apparent longitudinal magnetic field, we need
to bring the surface distribution of the longitudinal magnetic-flux density into an appropriate iso-radial velocity 
binning before we can add up their absolute contributions. 
\begin{table}[h]
\caption{Parameter ranges used for the model atmosphere and synthetic line calculations.} 
\centering                         
\begin{tabular}{c c c}    
\hline\hline       
Lower value & Parameter & Upper value \\    
\hline                       
   4500 K  & $T_{\rm{eff}}$  &  6500 K   \\     
   -2.0  & $[M/H]$  &  +0.2  \\
   2.0 & $\log g$  &  4.5  \\ 
   20 km/s  & v$\sin$$i$    &   75 km/s \\ 
   \hline                                   
\end{tabular}
\label{Table:1}
\end{table} 
The resolution of this iso-radial velocity binning 
depends on the width of the used (local) spectral line; i.e., the broader the spectral line, the less surface
flux can be deduced, and vice versa. 
Therefore we first translate the surface distribution of the longitudinal field of the model star 
onto a fine-binned rotational velocity coordinate axis. The resolution of this one-dimensional coordinate axis
has the same resolution as our synthetic spectral line profile, i.e. 0.5 km\,s$^{-1}$. The distribution of the 
longitudinal magnetic flux density along the rotation velocity is then convolved with an area-normalized
\emph{local} Stokes~$I$ profile to obtain the \emph{measurable} longitudinal flux-density spectrogram over the 
rotational velocity which is depicted in Fig.~\ref{Fig:5}.

We then sum over the absolute values of the longitudinal flux-density spectrogram to finally obtain the apparent
longitudinal magnetic field that is compared to the estimates of the OMP method. 
In our test case (phase=0.0), the true effective longitudinal magnetic field inferred from the model is 4.67 Gauss and the
apparent longitudinal magnetic field is 16.43 Gauss.
From the magnetic OMP algorithm, we obtain a value of 4.53 Gauss for the effective longitudinal magnetic field and 16.78 for the
apparent longitudinal magnetic field. 
An exhaustive statistical evaluation is given in the next section, but one can already see that besides 
the good approximation and the rapid convergence, the 
method also yields accurate values for both magnetic quantities.

\begin{figure}[!t]
\centering
\includegraphics[width=9cm]{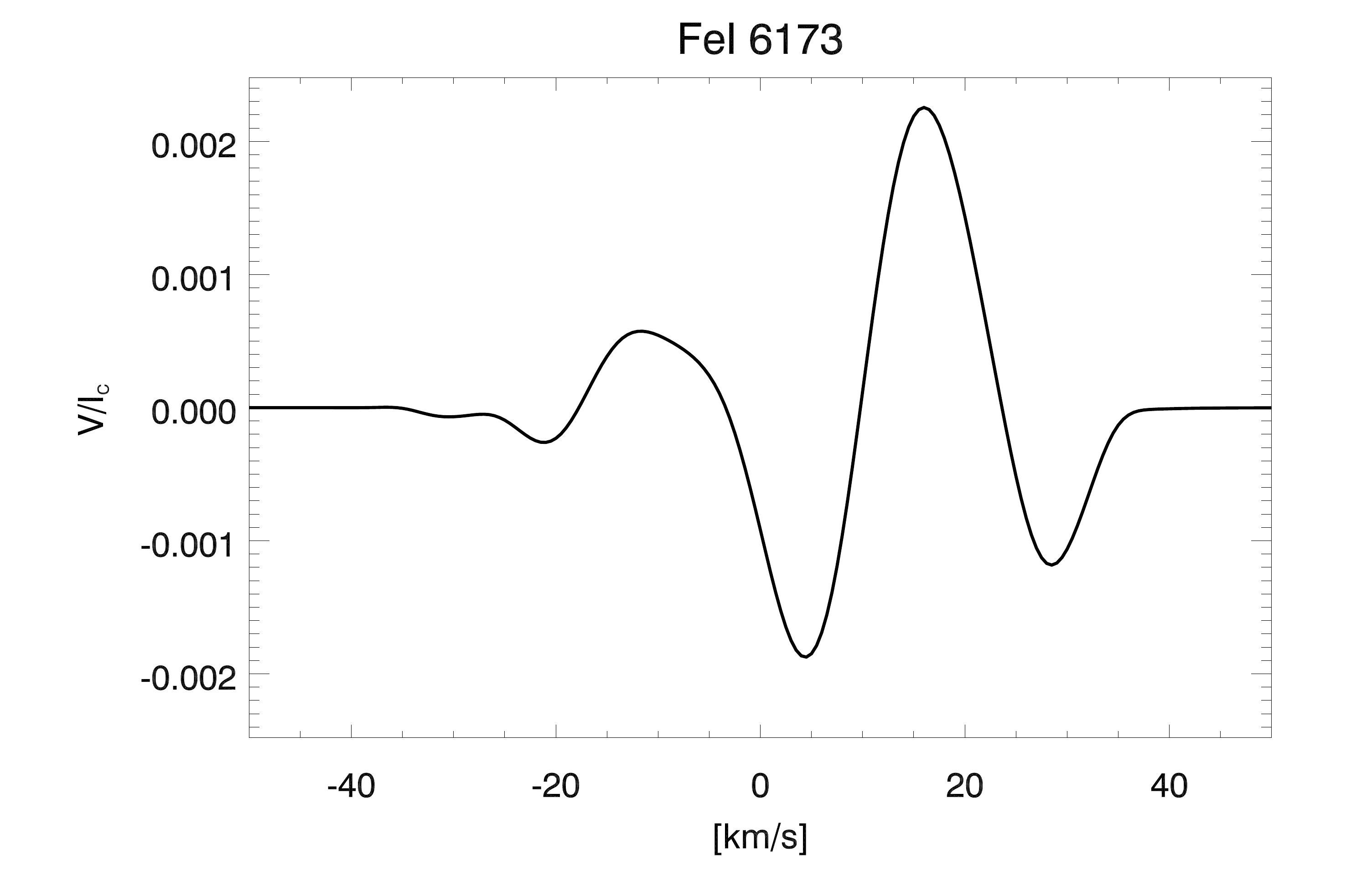}
\caption{Stokes~$V$ profile of the Zeeman sensitive iron line FeI 6173 for the
random distribution at phase 0.25 of the test star.} 
\label{Fig:6}
\end{figure}

There is another interesting effect that highlights the complementary nature of the
effective and apparent longitudinal magnetic field. At phase 0.25 of our test star, the polarities in the 
longitudinal magnetic field are almost perfectly balanced over the visible hemisphere. The resulting 
profile (again synthesized for the iron line FeI 6173) is shown in Fig.~\ref{Fig:6}. The almost perfect balancing of polarities
is not obvious from the line profile itself. However, if we estimate the longitudinal magnetic field by the COG method's 
Eq. (\ref{Eq:2.2.10}) and the magnetic OMP algorithm,
we obtain 0.005 G or 0.007 G, respectively, for the effective longitudinal magnetic field.

We thus have an extremely small longitudinal magnetic field or none at all. This is not what one 
would expect from the mere visual inspection of the Stokes~$V$ profile in Fig.~\ref{Fig:6}, 
which \emph{apparently} shows a clear net absolute polarization. In fact, the amplitude
of the Stokes~$V$ signal is even greater than the one of phase=0.0, shown in Fig.~\ref{Fig:2}.
This demonstrates the strength of the definition of the apparent longitudinal magnetic field; 
the magnetic OMP algorithm detects a clear apparent longitudinal field of 18.63 G (true value 17.98 G), 
which is slightly stronger even than in phase 0.0.
The quantity of the apparent magnetic longitudinal field is therefore of particular interest in cases where 
balanced field distributions cause the first-order moment of the Stokes~$V$ profile to vanish. 
\begin{figure}[!t]
\centering
\includegraphics[width=9cm]{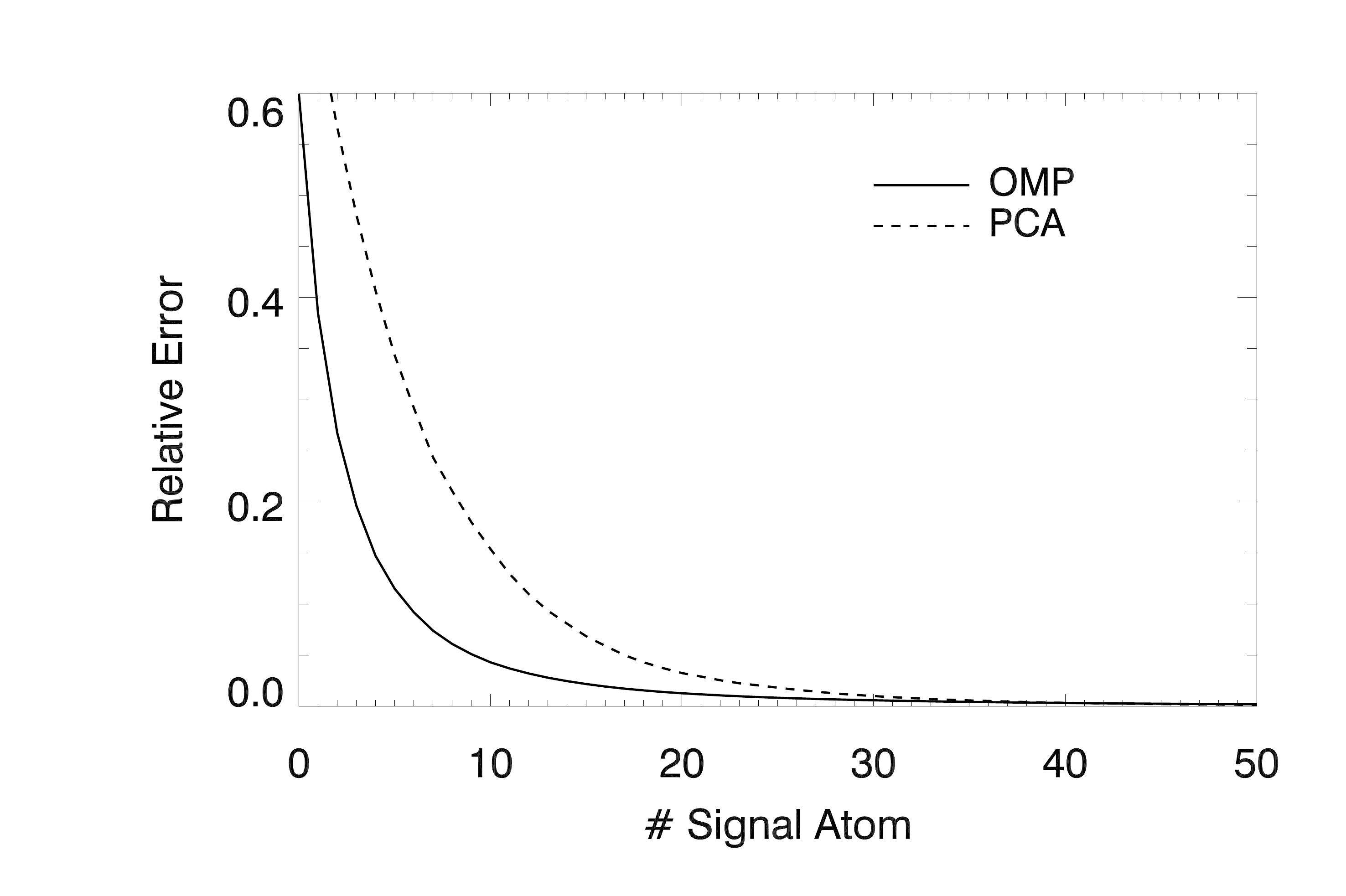}
\caption{The mean relative approximation error for the OMP dictionary and the PCA expansion. The relative error of
the approximation is plotted over the number of signal atom (eigenprofiles for the PCA). The error declines steeply 
for both expansion methods; however, the OMP expansion shows a better performance, i.e., sparsity behavior compared to the PCA.}
\label{Fig:7}
\end{figure}
\section{Numerical experiments and statistical evaluation}  
\label{Sect:5}
In this section we assess the accuracy of the OMP algorithm under a broad range of model conditions
and for different spectral lines.
For that reason we synthesized a large number of Stokes~$I$ and Stokes~$V$ profiles for different magnetic 
surface distributions and atmospheric conditions. 
The model parameters are the effective temperature, metallicity, surface gravity, 
projected rotational velocity, and the surface magnetic field and its distribution. 
For each set of the atmospheric parameters, we created a random surface distribution
in the same way as for the previous example in Fig.~\ref{Fig:1}.  
The atmospheric parameters 
were randomly chosen from within an interval given in Table \ref{Table:1}. For the model atmospheres, we chose 
Kurucz/Atlas-9 models \citep{Castelli04}. 
\begin{figure*}[!t]
\begin{minipage}{\textwidth}
\centering
\includegraphics[width=9cm]{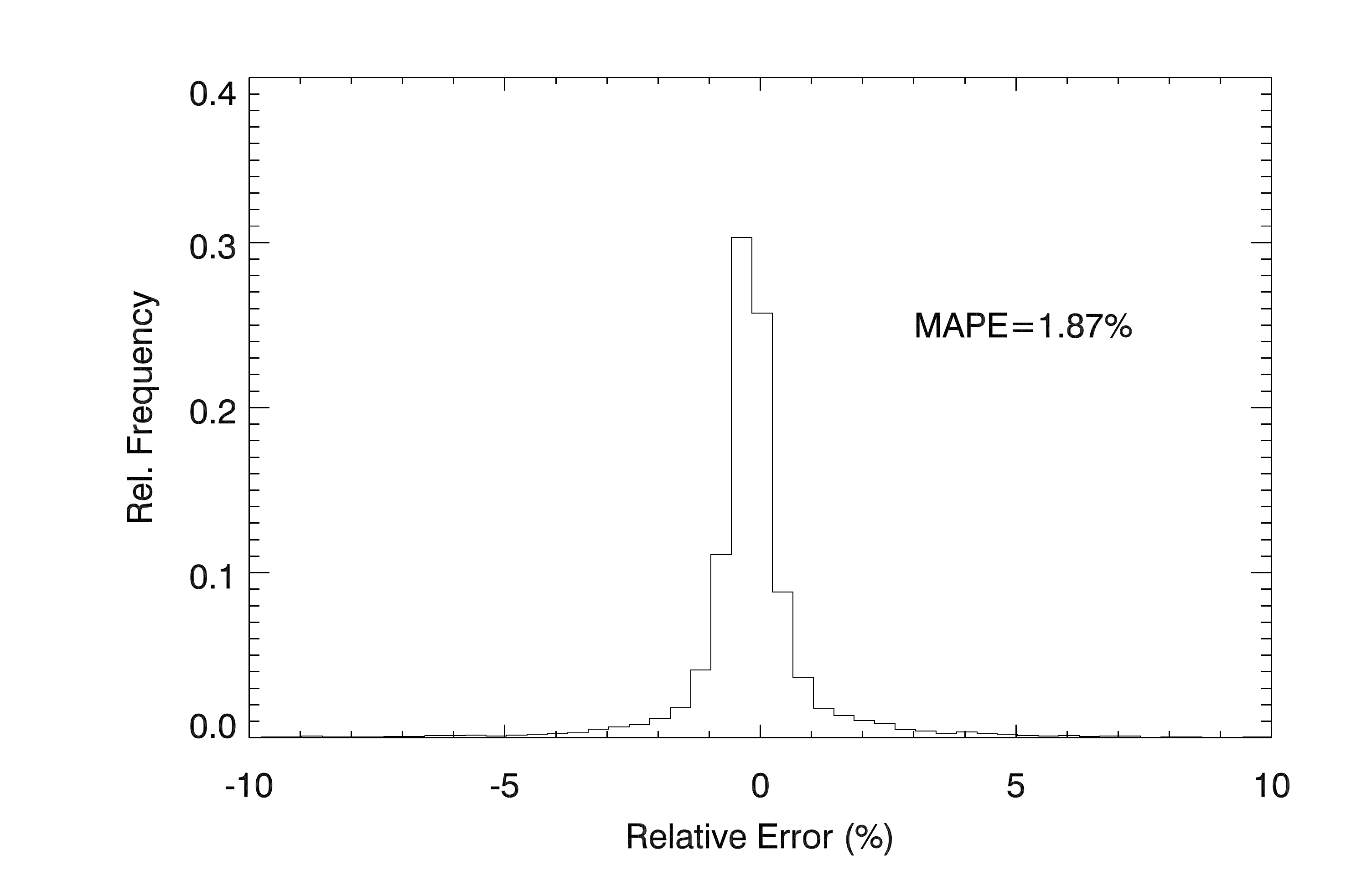}
\includegraphics[width=9cm]{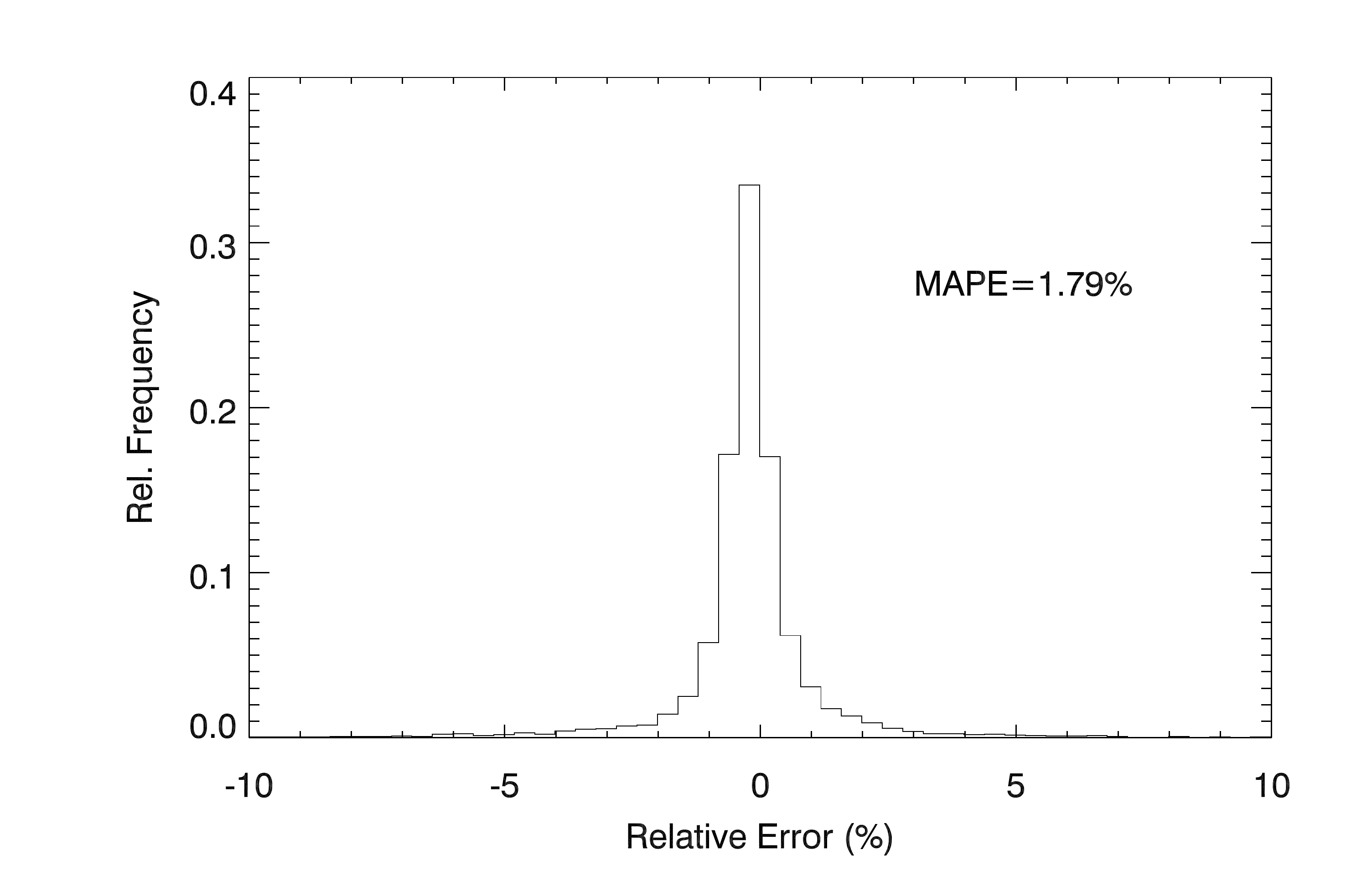}
\end{minipage}
\caption{Error distribution of the longitudinal magnetic field calculated from 15,000 synthetic Stokes profiles by the 
magnetic OMP method (left) and the center-of-gravity (COG) method (right). 
Both methods show very good accuracy with a mean absolute percentage error of 1.87~\% (magnetic OMP) 
and 1.79~\% (COG).}
\label{Fig:8}
\end{figure*}
Because the model atmospheres are provided on a fixed grid 
of step sizes of 250 K for the effective temperature, 0.5 for the logarithmic abundance and gravity,
the atmospheres are interpolated for each randomly chosen set of atmospheric parameters.
As Zeeman-sensitive spectral lines we chose three magnetically sensitive lines; 
the iron lines FeI 5497 ($g_{\rm eff} = 2.22$), 
FeI 6173 ($g_{\rm eff} = 2.50$), and FeI 8468 ($g_{eff} = 2.50$). All line parameters were again taken 
from the VALD line database \citep{Piskunov95,Kupka99}. For each spectral line, we created a sample
of 5,000 randomly chosen atmospheric parameters, rotational velocities, and magnetic surface distributions 
to eventually synthesize a set of corresponding Stokes~$I$ and Stokes~$V$ profiles with the forward module of 
our \emph{iMap} code. 

In total, we calculated a set of 15,000 Stokes~$I$ and Stokes~$V$ profile, which were then analyzed 
by our magnetic OMP algorithm.
The true effective longitudinal magnetic field is directly extracted from the magnetic surface
distribution of the synthetic model star.
The true apparent longitudinal magnetic field is obtained 
from the convolved magnetic spectrograms as described above.
Because the individual trials, sets of atmospheric parameters, and surface magnetic field values have 
different scales, we chose a relative performance metric to describe the accuracy. 

The overall performance is judged by the mean absolute percentage error, while for illustrating the 
error distribution we use the relative percentage error. The relative percentage error is 
defined as
\begin{equation} 
D_i \: = \: \frac{B^T_i - B^{OMP}_i}{B^T_i} \: * \: 100 \: ,
\end{equation}
where $B^T_i$ is the true magnetic field, and $B^{OMP}_i$ the calculated magnetic field 
of the $i$-th sample. Finally, the mean absolute percentage error (MAPE) is defined by
\begin{equation}
 M_B \: = \: \frac{1}{n} \: \sum_{i=1}^n \: \left | D_i \right | \: .
\end{equation}

\subsection{Sparsity of Stokes profiles}
\label{Sect:5.0}
Before we evaluate the performance of the OMP algorithm, we look more closely at the general sparsity 
of Stokes profiles. A sparse representation is often achieved by transforming a vector from the 
original data domain into a domain where the transformed coefficients (i.e., expansion coefficients) 
allow a more compact representation of the information (e.g., Fourier or wavelet transform).  
A vector is called sparse if there is a representation where most of the vector entries are zero or close to zero.
A sparse approximation is performed by restricting the sparse transformation to the largest expansion coefficients
such that there is no great loss of signal information. 

Recently, \citet{Asensio10} have shown that Stokes profiles are 
sparse and compressible under various transformations (e.g., wavelets and empirical basis functions). 
Compressible here means that the values of the sorted expansion coefficients exhibit an exponential 
decay that in turn results in a small approximation error \citep{Candes08}.
Although we saw from the numerical simulation of the last section that a sparse representation of a synthetic Stokes profile
is possible with our dictionary elements, we also want to test empirically if this also holds for the entire set of our 
15,000 Stokes profiles. To quantify the approximation error, we use the relative error, 
$\sum_i^n \frac{1}{n} \frac{\|\hat{\vec{S}}_i - \vec{S}^*_i \|}{\| \vec{S}^*_i \|}$ 
where $\hat{\vec{S}}$ is the approximation and $\vec{S}^*$ the 
original Stokes vector. 

To compare the performance of the OMP expansion relative to a known sparse decomposition, we also calculated a 
principal component analysis (PCA) of the entire synthetic data set. A PCA decomposes a set of observed Stokes profiles
into an empirical set of orthogonal eigenprofiles \citep[e.g.,][]{Carroll08}.
As has been shown by \citet{Asensio10} and \citet{Asensio10b}, a PCA expansion can result in a compact and sparse representation of Stokes profiles.
Figure \ref{Fig:7} shows how the dictionary used with the OMP algorithm performs against a PCA decomposition of our synthetic data set.
Both curves (solid OMP, dashed PCA) show a rapid decline of the mean approximation error 
with increasing numbers of signal atoms or eigenvectors, respectively. 
\begin{figure}
\centering
\includegraphics[width=9cm]{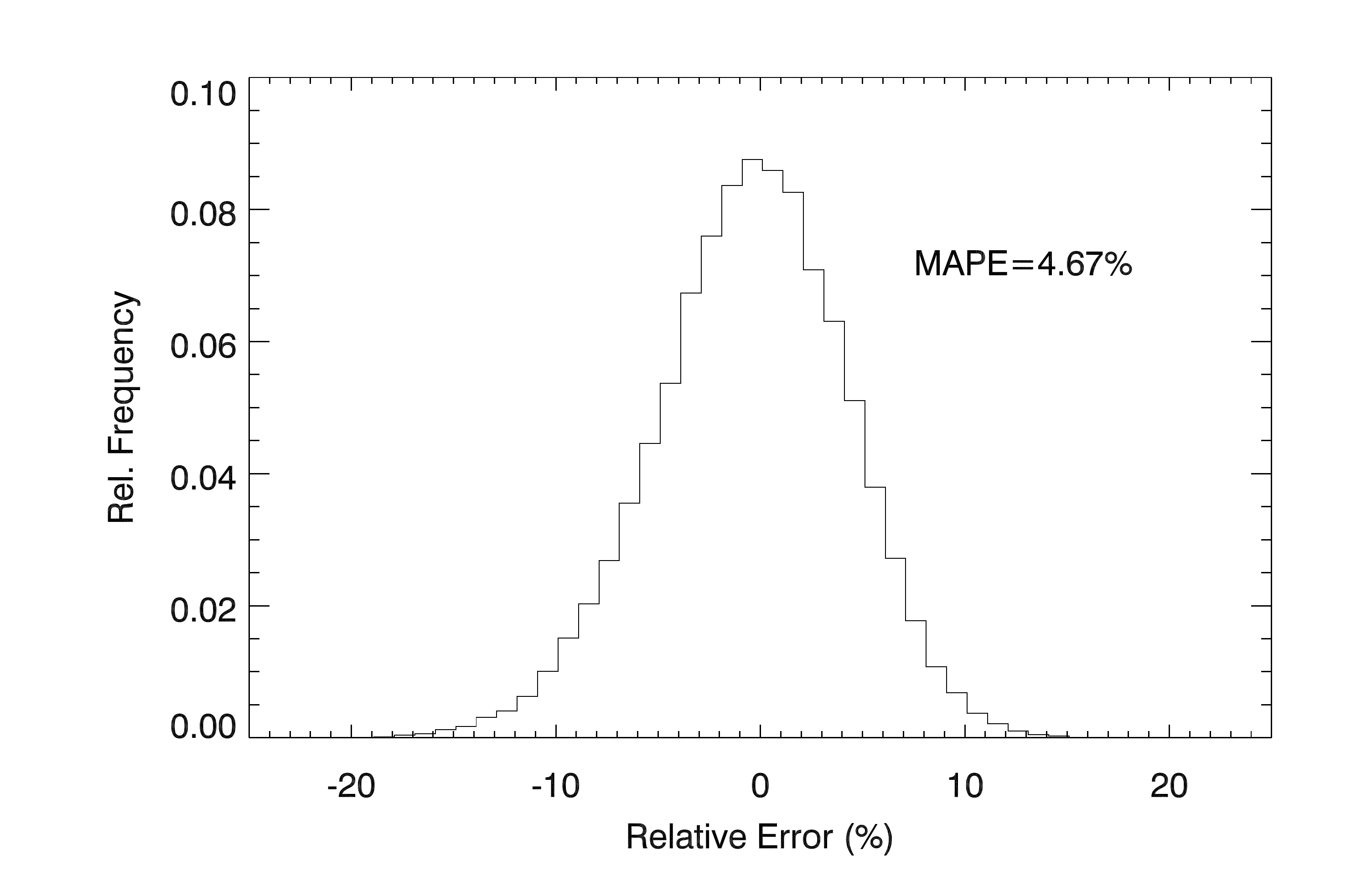}
\caption{Error distribution of the \emph{apparent} longitudinal magnetic field calculated by the magnetic OMP method. The mean absolute percentage
error from the entire set of 15,000 synthetic test calculations is 4.67~\%.} 
\label{Fig:9}
\end{figure}
The performance of our dictionary of Gaussian derivatives is better than that of the empirical basis functions (i.e., eigenprofiles)
obtained by the PCA. The reason for that is two-fold: First, the signal atoms of our dictionary are selected a priori to match the 
building blocks (i.e., elementary waveforms) of our problem and second the redundancy of our dictionary facilitates a more efficient 
expansion then that of the orthogonal eigenvectors of the PCA.

In contrast to the empirical basis formed by the PCA, the OMP
dictionary is not required to be a set of orthogonal basis functions, and this generally provides better 
flexibility and adaptivity to approximate a given signal \citep{Candes08}.

To reduce the relative approximation error with the OMP algorithm below 5\%, it 
takes on average only nine signal atoms. Using 22 signal atoms the relative error is on average smaller than 1 \%, 
which demonstrates that our dictionary allows a sparse representation of Stokes profiles for the
parameter regime given in Table \ref{Table:1}.
\begin{figure}[!t]
\centering
\includegraphics[width=9cm]{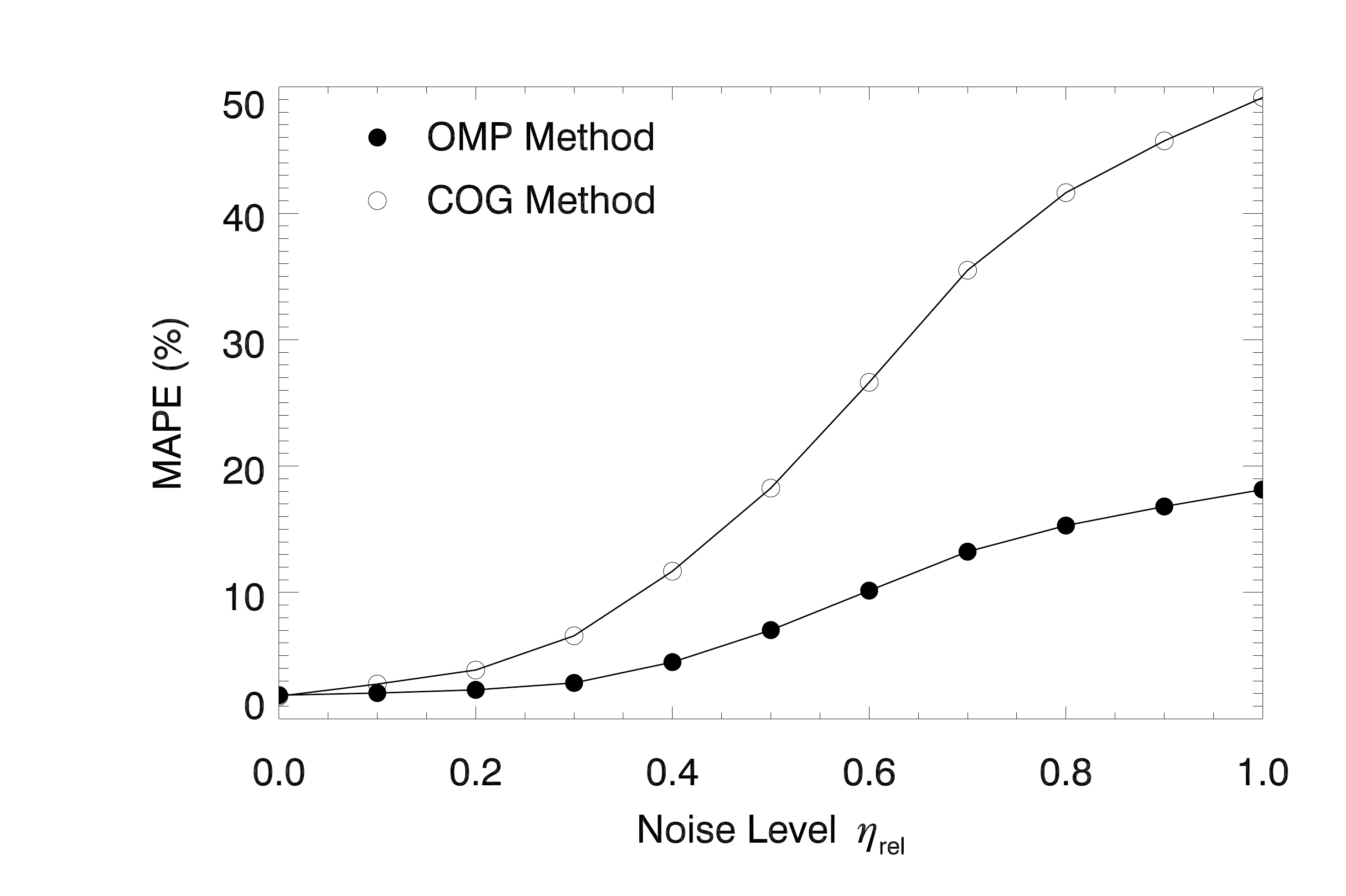}
\caption{Mean absolute percentage  error (MAPE) of the effective longitudinal magnetic field calculated over the relative noise level for the 
COG method (open circles) and the magnetic OMP method (solid circles).} 
\label{Fig:10}
\end{figure}
\begin{figure}[t]
\centering
\includegraphics[width=9cm]{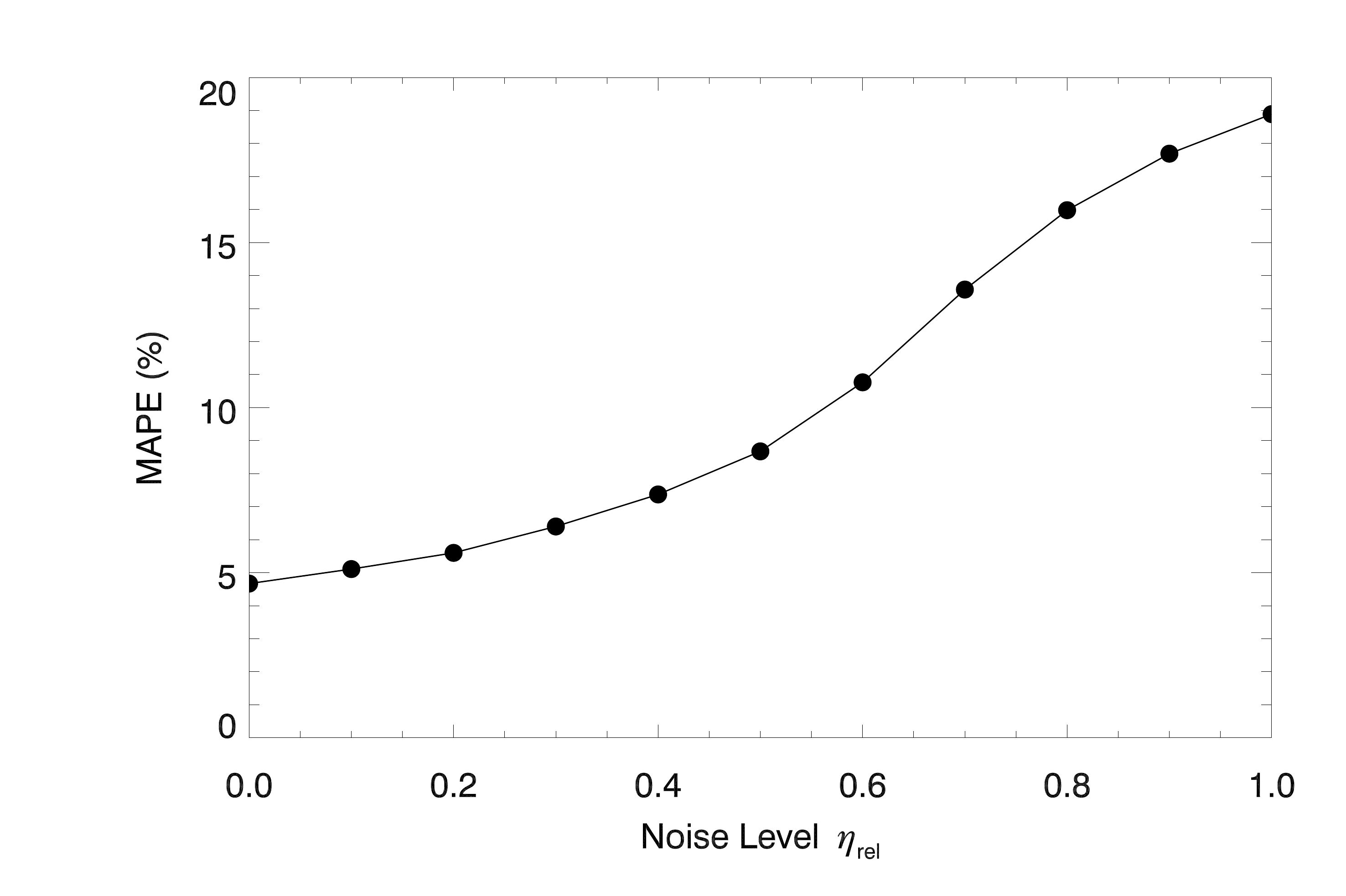}
\caption{Mean absolute percentage error (MAPE) of the apparent longitudinal magnetic field calculated versus the relative noise level.} 
\label{Fig:11}
\end{figure}
\begin{figure}[t]
\centering
\includegraphics[width=9cm]{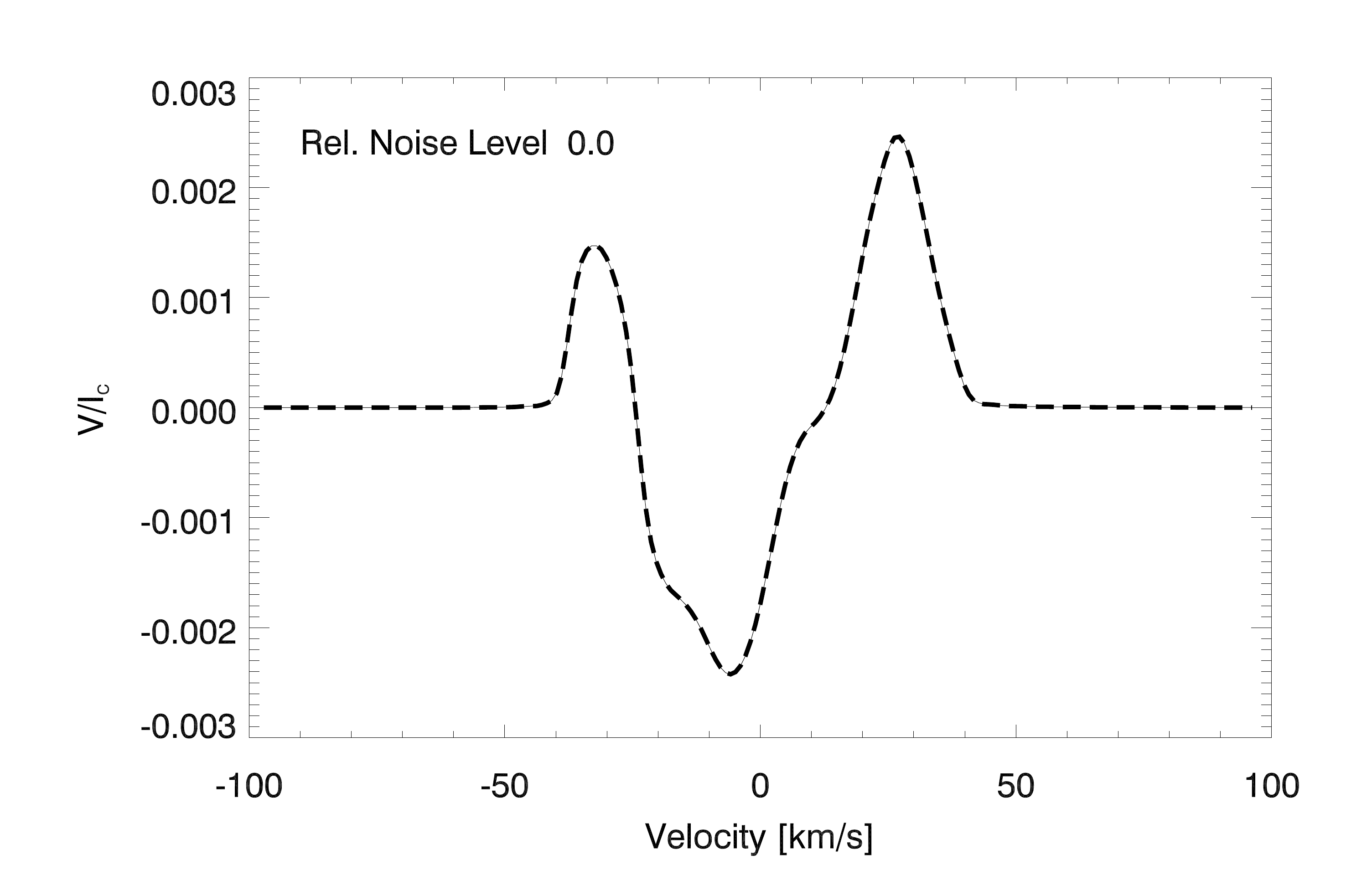}
\caption{A noiseless (noise level 0.0) example profile (solid line), and its OMP approximation (dashed line). } 
\label{Fig:12}
\end{figure}
      
\subsection{The noise-free case}
\label{Sect:5.1}
To obtain an idea about the principal accuracy of the magnetic OMP algorithm and its performance relative to the 
conventional COG method, we began with a noise-free test case where the entire test sample of 15,000 
Stokes~$I,$ and Stokes~$V$ profiles were used without any noise contribution.

In this simulation the overall error value for the effective longitudinal magnetic field obtained from the magnetic OMP algorithm yielded a 
MAPE of 1.87~\%. For comparison the effective longitudinal magnetic field calculated by the COG method shows a similar 
MAPE of 1.79~\%. The same error for the apparent longitudinal magnetic field is 4.67~\%. 
The error for the three individual spectral lines show no apparent deviation from the overall error. 

For the subset of synthetic Stokes~$V$ 
profiles of the FeI 5497, we obtain a MAPE for the effective (apparent) longitudinal magnetic field of 1.80~\% (4.57~\%), 
for the FeI 6173 lines a MAPE of 1.93~\% (4.69~\%), and for the FeI 8468 lines a MAPE of 1.88~\% (4.72~\%).
The error distribution for the relative percentage error of the effective longitudinal field calculated by the OMP and the COG method
is shown in Fig.~\ref{Fig:8}. Both the OMP method and the conventional COG method provide equally good results.
The error distribution of the apparent longitudinal magnetic field is shown in Fig.~\ref{Fig:9}. Given that the apparent field 
is even harder to determine due to possible cancellation effects of the Stokes~$V$ signals, the error is still remarkably low, and the OMP method 
is able to recover the apparent field from the Stokes~$V$ profiles with good accuracy.

\subsection{The noise case}
\label{Sect:5.2}
In the following we test the accuracy of the magnetic OMP method and the COG method for the more relevant case when the Stokes~$V$ profiles
are contaminated with noise. For that reason we have added an increasing amount of white noise to the 15,000 Stokes~$V$ profiles. 
Since the Stokes~$V$ profiles generated from the random surface distribution are of different magnitudes, we define a relative 
noise level $\eta$ that is given by the relative variance of the Stokes~$V$ profile and that of the noise according to 
\begin{equation}
\eta_{\rm rel} \: = \: \frac{\sigma(N)}{\sigma(S)} \: ,
\end{equation}
where $\sigma(S)$ and $\sigma(N)$ are the standard deviations of the Stokes~$V$ profile and the noise contribution, respectively.
Figure \ref{Fig:10} shows how the relative error increases with higher noise levels. It is interesting how rapidly the accuracy  
of the COG method deteriorates (upper curve) compared to the relative robustness of the magnetic OMP method 
(lower curve). Already for a noise level of $\eta_{\rm rel}$ = 0.3, the accuracy of the COG method 
is significantly affected, whereas the OMP method still provides good accuracy with errors of less than 10~\%. 
Even for a noise level of unity, the OMP method has an error of just 18~\% where the 
COG method with an error of almost 50~\% can no longer be considered a meaningful tool for quantifying the magnetic field.
For the apparent longitudinal magnetic field, the OMP method shows a very robust performance as well. The error distribution
over the relative noise level is illustrated in Fig.~\ref{Fig:11}.
The good performance and robustness of the magnetic OMP method compared to the COG method can be understood 
by considering the noise-free Stokes~$V$ profile in Fig.~\ref{Fig:11}.
Because the OMP algorithm gradually approximates the noisy profile with a linear combination of dictionary 
profile atoms, it calculates the effective and apparent longitudinal magnetic field 
for each of the selected profile atoms separately. The detection mechanisms of Sect.\ref{Sect:3.3} 
prevent the algorithm from approximating insignificant profile features and thus largely avoids the 
quantification of the contributing noise.
\begin{figure*}[t]
\begin{minipage}{\textwidth}
\centering
\includegraphics[width=9cm]{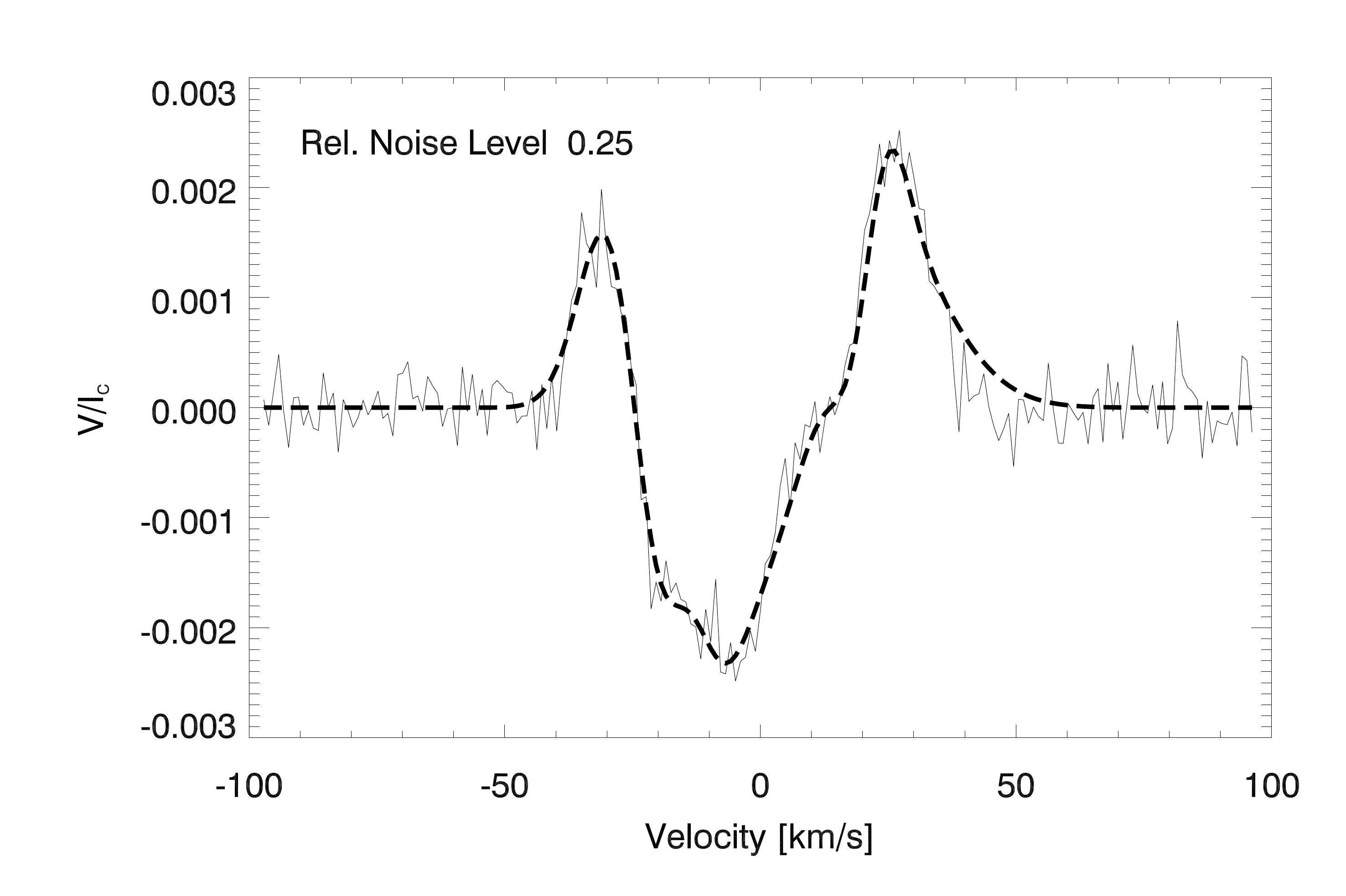}
\includegraphics[width=9cm]{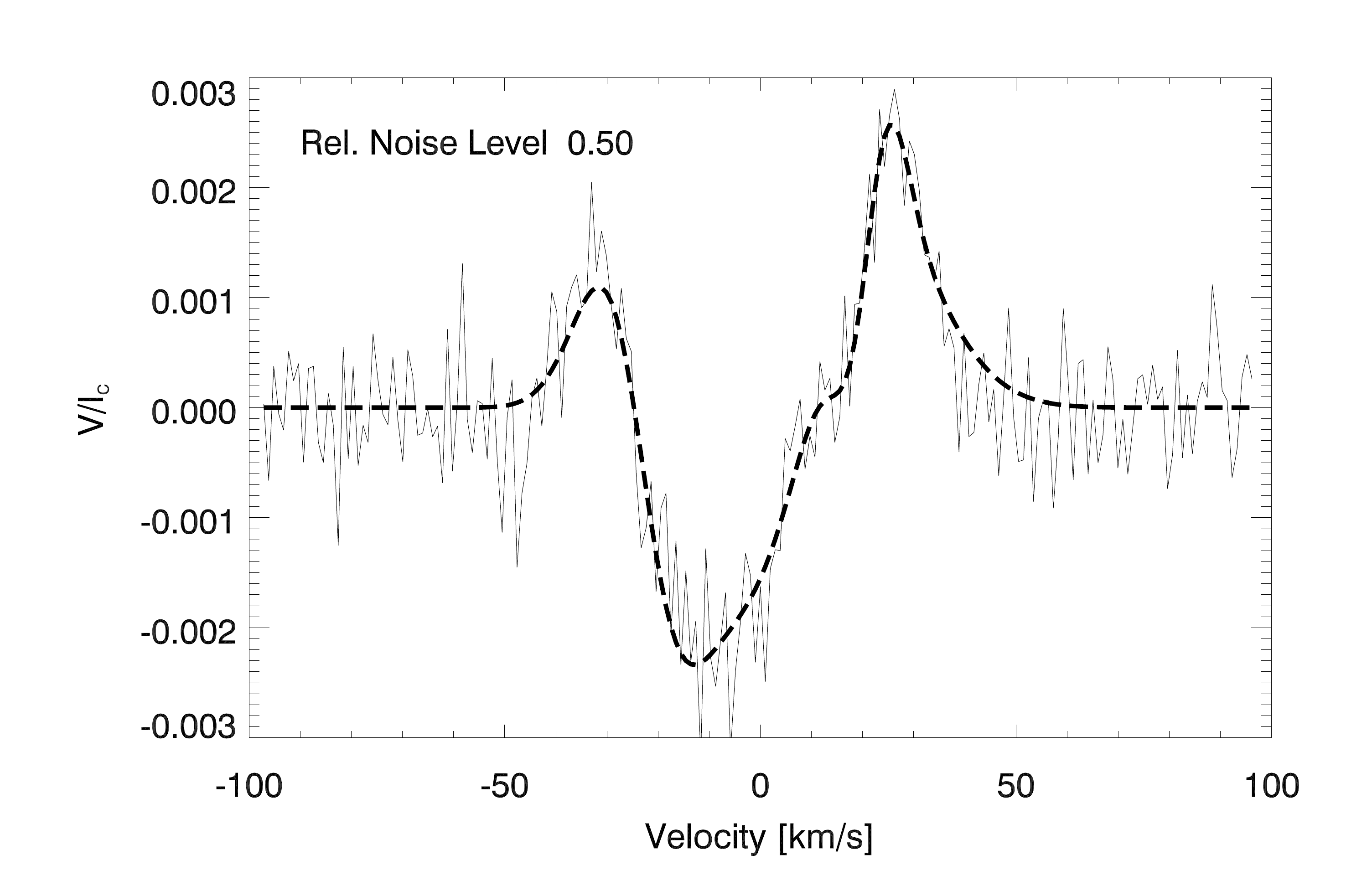}
\includegraphics[width=9cm]{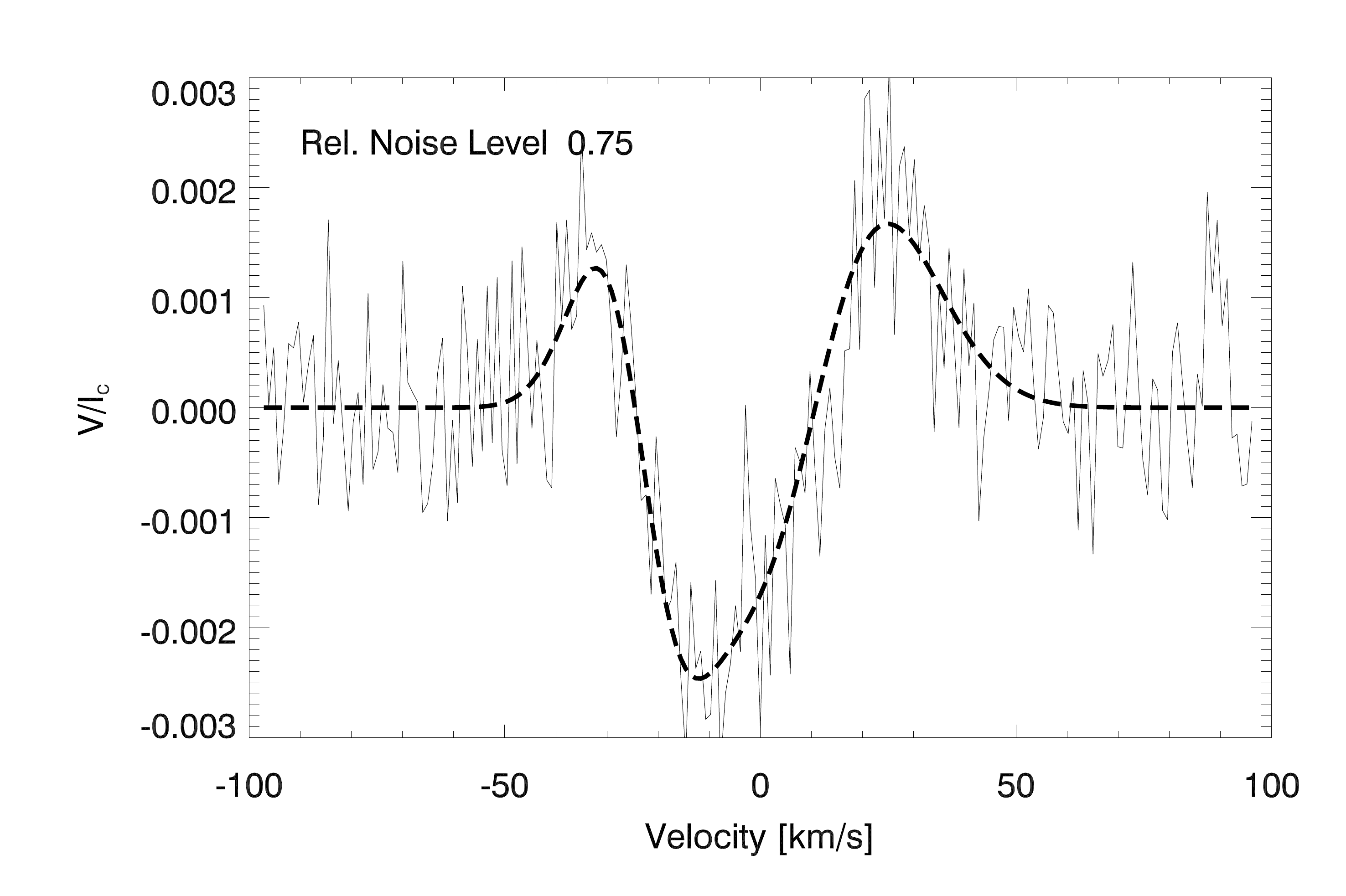}
\includegraphics[width=9cm]{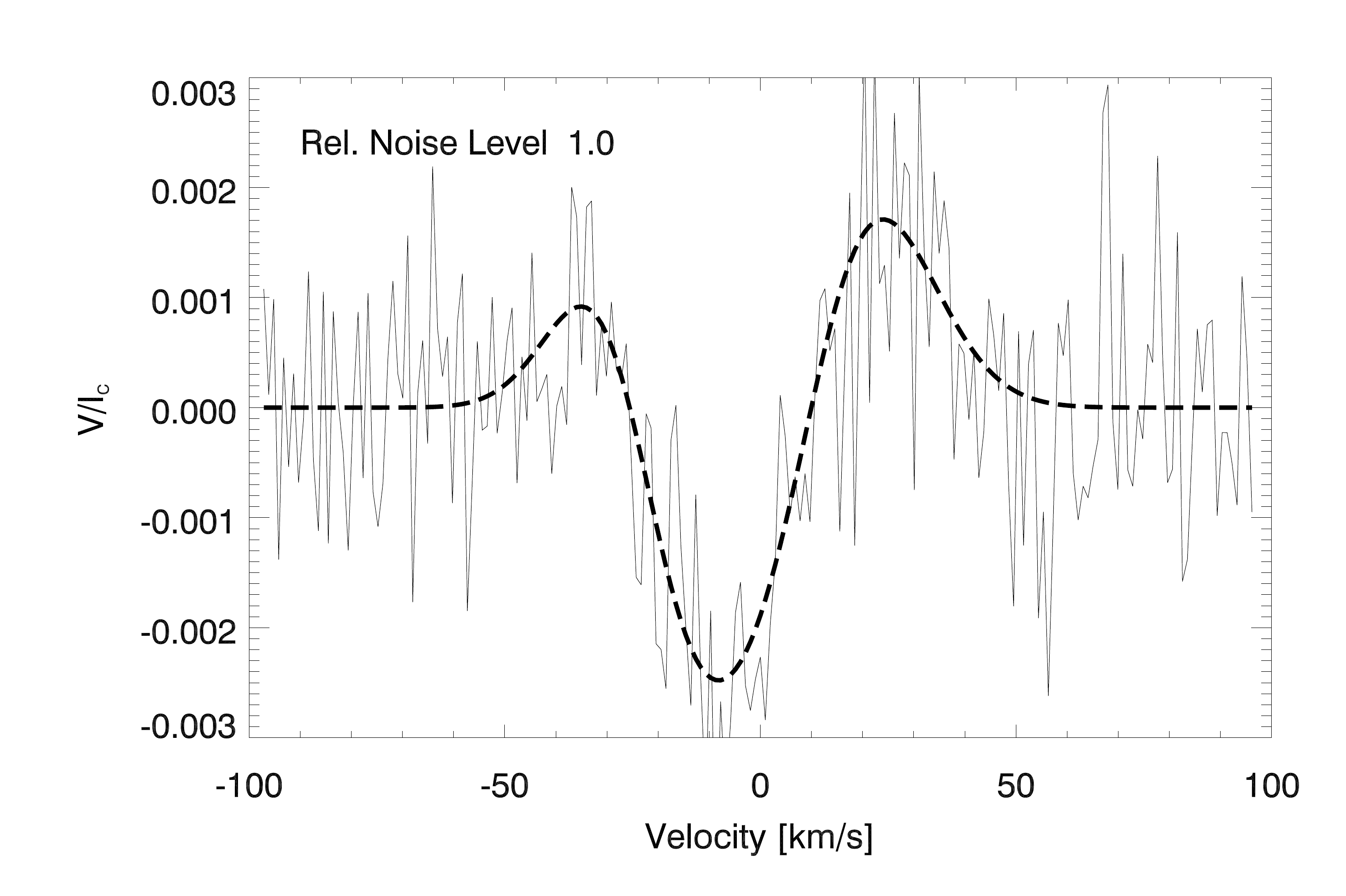}
\end{minipage}
\caption{Magnetic OMP approximations (dashed lines) to a noisy Stokes~$V$ profile (solid lines) 
for different relative noise levels. Despite the increasing relative noise, the OMP approximation 
identifies the main characteristic and shape of the underlying noise-free profile.}
\label{Fig:13}
\end{figure*} 
Figure \ref{Fig:12} shows one of the sample profiles generated for a random surface distribution of the
magnetic field. The noiseless profile is approximated well by the OMP method, 
and the estimated effective and apparent longitudinal magnetic field is approximated with --41.04 G and 59.88 G, 
respectively, which is very close to the true values of --40.11 G and 61.65 G.
The noiseless profile is approximated by the maximum number of 40 iterations, which means that 40   
dictionary profile atoms are used to approximate the example profile. However, already the first 
three profile atoms are able to approximate the profile to such a degree that 95 \% of the signal variance are 
described. 
In Fig.~\ref{Fig:13} the same profile can be seen with varying degrees of noise contributions overplotted
again by the approximation from the OMP algorithm.
Despite the increasing relative noise level, the OMP algorithm finds a good approximation to the original Stokes~$V$ profile. 
The iteration for a noise level of $\eta_{\rm rel}$ = 0.25 stops already at seven iterations (i.e., 7 profile atoms). 
For $\eta_{\rm rel}$ = 0.5, 0.75, and 1.0, the iteration is stopped at cycles 6, 4, and 3, respectively. At a relative noise level of unity only 
three signal atoms are used to approximate the Stokes~$V$ profile. However, as one can see from the noiseless case, these three signal atoms 
already capture the essential profile shape and amplitude of the original Stokes~$V$ signal. 
Since the magnetic OMP algorithm interprets and analyzes each signal atom separately, the good approximation directly translates 
to a good estimates of the underlying effective and apparent longitudinal magnetic field.  
In this particular case, the estimations for the effective (apparent) longitudinal field 
from the magnetic OMP algorithm is --42.54 G (64.27 G) for a noise level of 0.25, 
--44.26 G (65.71 G) for a noise level 0.5, --36.04 G (68.85 G) for a noise level 0.75, and --34.77 G (72.54 G) for
noise level of 1.0.

\subsection{Absolute limits of the COG and OMP method}
\label{Sect:5.3}
Until now we have analyzed the performance of the magnetic OMP and COG method
by using \emph{relative} noise levels. To shed more light on the absolute performance limits of the two methods, 
we take a closer look on the absolute values of the noise. Multiline reconstruction techniques like 
LSD \citep{Donati97} or SVD \citep{Carroll12} allow us to drastically lower the noise levels of spectropolarimetric observations.
The thus increased signal-to-noise (S/N) levels allows detecting faint signal profiles that were otherwise 
deeply buried in the noise. 

A natural question that arises in the context of magnetic field estimation is how much of the 
residual noise is falsely interpreted as magnetic field and up to which limits 
we can reliably estimate a true underlying magnetic field.
To address this problem, we could simply perform a first-order perturbation to the COG method and apply standard error propagation
to estimate the influence of the noise. However, for weak magnetic fields, the noise contribution relative to the true 
signal amplitude is in general not small anymore. 
We therefore simulate the process with a large and statistically significant number of synthetic profiles to determine the impact of typical 
absolute noise levels on the magnetic field estimation. 

We assume additive Gaussian noise with zero mean. One 
might expect that this would lead to a vanishing value 
in the field estimation by the COG method. This is, however, not true for a finite spectral resolution.
The random fluctuations and the unequal weighting in the velocity (or wavelength) domain can lead to relatively large 
spurious contributions to the first-order moment, hence to the field estimation as shown in the following simulation.
Note that any form of correlated noise would even exacerbate the problem since it would introduce a systematic bias
in the evaluation of the first-order moment.

With our assumption of white noise, a Stokes~$\tilde{V}^*$ profile can be written as the sum of the true noiseless Stokes~$\tilde{V}$ profile 
and a noise vector $N$, 
\begin{equation}
\tilde{V}^*(\textrm{\textsl{v}}) \: = \: \tilde{V}(\textrm{\textsl{v}}) \: + \: N(\textrm{\textsl{v}}) \: .
\end{equation}
The effective or mean longitudinal field as measured by the COG method can then be written as 
\begin{equation}
B_{\rm eff} = \frac{\lambda_0 \: \int_{V} \left ( \tilde{V}(\textrm{\textsl{v}}) + N(\textrm{\textsl{v}}) \right ) \: 
\textrm{\textsl{v}} \: d\textrm{\textsl{v}}} {c \: \alpha  \: W_I} \: .
\end{equation}
We assume that the noise level is small enough that the estimation of the Stokes~$I$ equivalent width remains largely unaffected.

The first-order moment of a noisy Stokes~$\tilde{V}^*$ profile, as well as the magnetic field estimation, can then be 
split into a signal ($B^S_{\rm eff}$) and a noise ($B^N_{\rm eff}$) contributing part,
\begin{equation}
B_{\rm eff} \: = \: B^S_{\rm eff} \: + \: B^N_{\rm eff} \: = \: \frac{\lambda_0} {c \: \alpha  \: W_I} \: \left ( \int_{V} \tilde{V}(\textrm{\textsl{v}}) \: \textrm{\textsl{v}} \: d\textrm{\textsl{v}}
+ \int_{V} N(\textrm{\textsl{v}}) \: \textrm{\textsl{v}} \: d\textrm{\textsl{v}}  \right ) \: .
\label{Eq:5.3.1}
\end{equation}
Written in this form we may ask to which noise level the 
following relation is valid, $ B^S_{\rm eff} > B^N_{\rm eff}$, or at which point does a noise-induced 
magnetic field interfere with the magnetic field estimation from the true signal profile. 

For the simulations we defined the absolute noise level $\eta_{abs}$ as the standard
noise deviation $\sigma_N$ 
relative to the continuum of the normalized intensity profile, i.e., $\eta_{\rm abs} = \sigma_N,$ which is also commonly
used to describe the S/N level ($1/\sigma_N $) in stellar polarimetry \citep{Donati97}. To mimic a multiline,
reconstructed line profile (e.g., LSD profile), we created a fictitious iron line with a mean Land\'e factor of 1.2 
at a rest wavelength of 5000~\AA. The oscillator strength and van der Waals damping were adjusted to give an 
equivalent width for the Stokes~$I$ profile of 100~m\AA. 
The Stokes~$I$ and Stokes $V$ noise profiles were calculated with a resolution of $\lambda/\Delta\lambda=$100,000, which is
common to current high-resolution observations. The spectral range covers a velocity of plus/minus 50 km\,s$^{-1}$ around 
the line center, and integration limits were set accordingly. 

We chose noise levels from 10$^{-3}$  down to 10$^{-5}$ that correspond to a polarimetric sensitivity of
0.001 \%, a level that is reached by current reconstruction methods \citep{Donati09}.
We selected 20 noise levels between 10$^{-5}$  and 10$^{-3}$, equally spaced on a logarithmic scale. 
For \emph{each} absolute noise level, we synthesized 10~000 noisy line profiles such that we have a total
set of 200~000 simulated profiles. 
\begin{figure*}[!t]
\begin{minipage}{\textwidth}
\centering
\includegraphics[width=9cm]{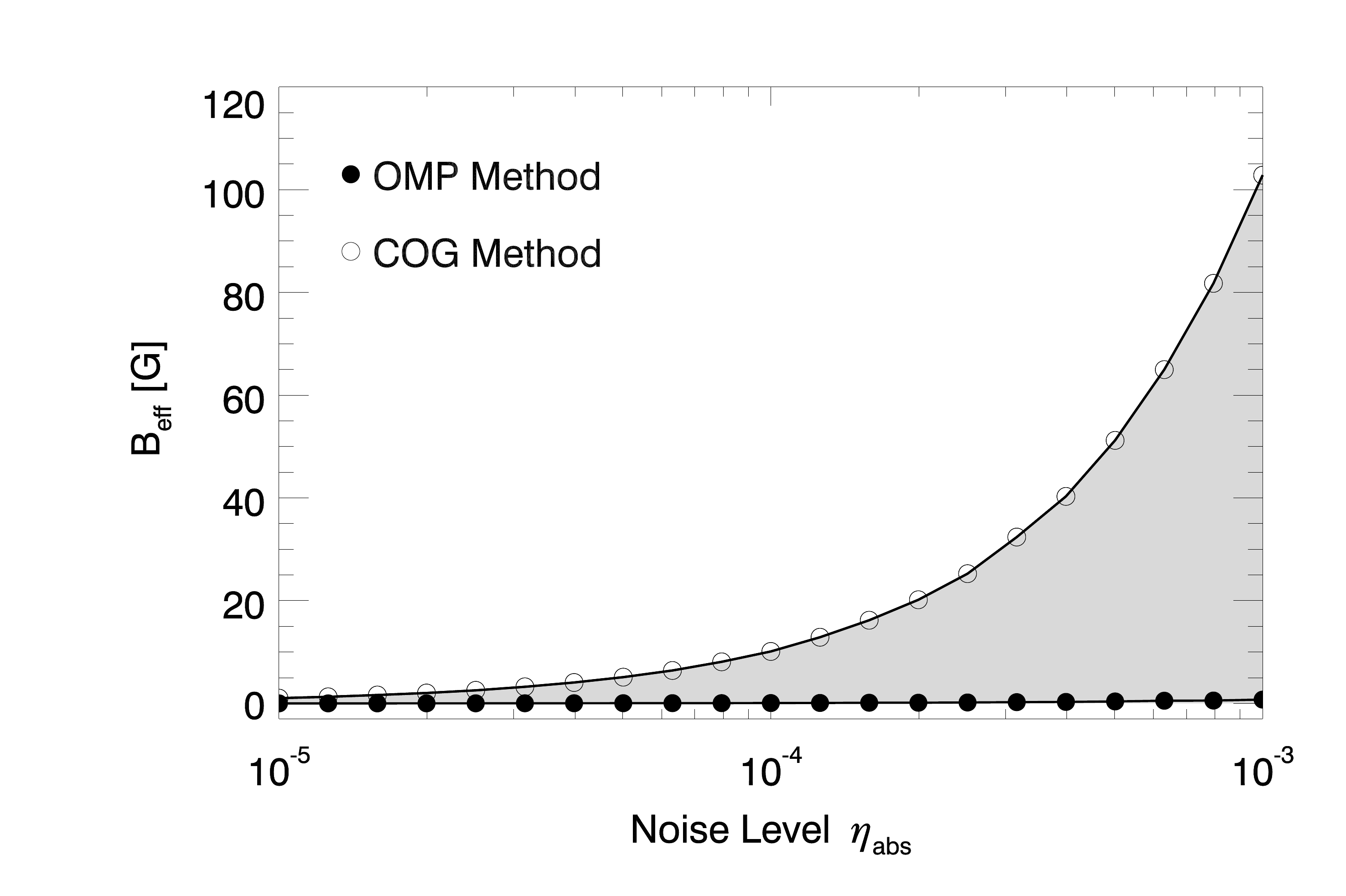}
\includegraphics[width=9cm]{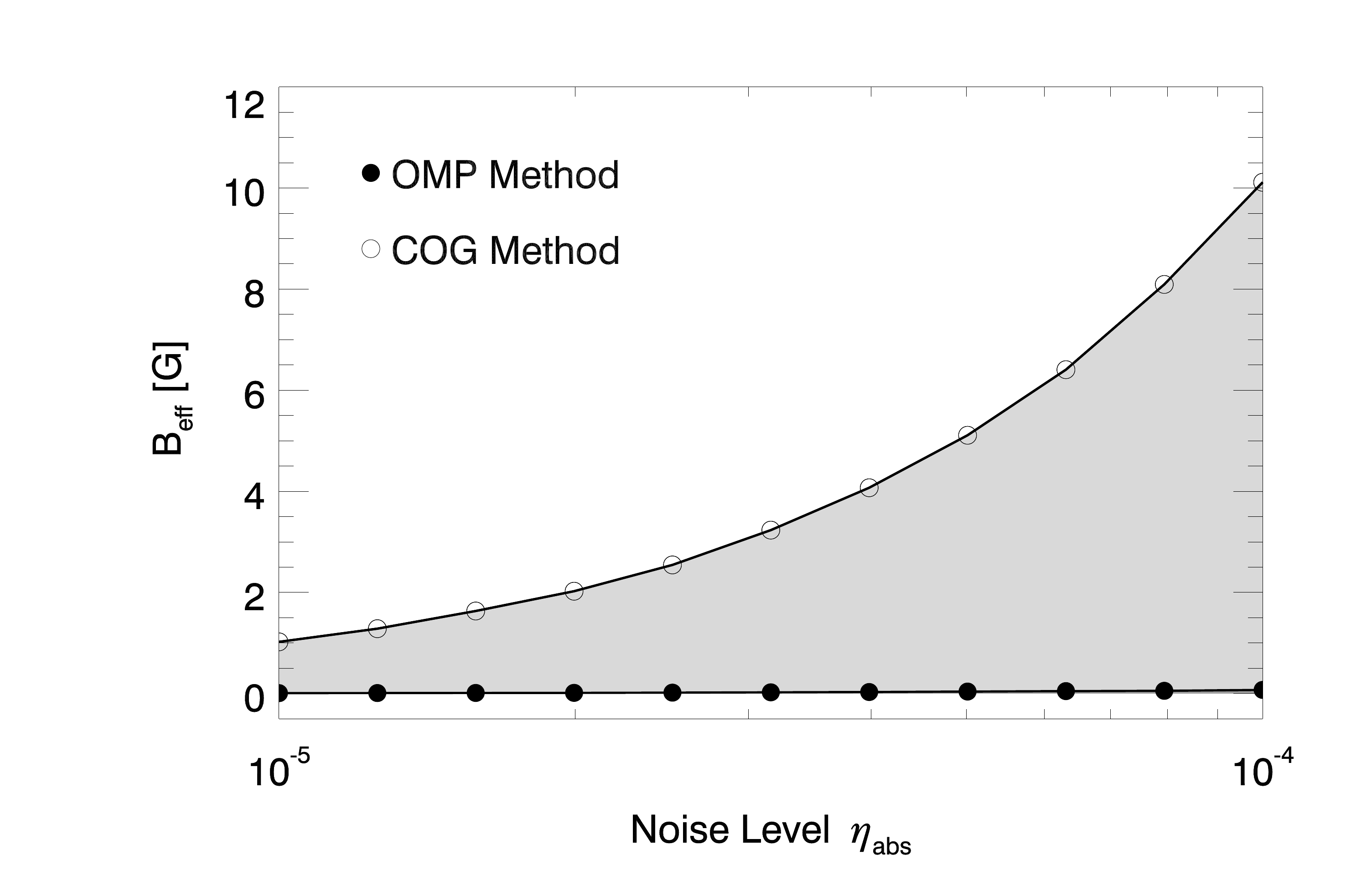}
\caption{The effective magnetic field response over the absolute noise level (logarithmic scale) 
for the COG method (open circles) and the OMP method (solid circles). 
The response curve shows how much of the noise is erroneously attributed to a magnetic field. 
On the left side, the response is shown for a noise range between 10$^{-5}$ and 10$^{-3}$. The noise response 
of the COG method rapidly increases with the noise level from 2 to 100~G, whereas the OMP method 
remains largely unaffected by the noise with a flat response not larger than 0.7~G. 
On the right side, a close-up of the magnetic response curve for the region between 10$^{-5}$ to 10$^{-4}$ is shown. 
Even in such a low noise regime, the COG method falsely interprets a significant portion of the noise as 
magnetic field. The OMP method stays essentially flat at a level close to zero over the entire region.
The gray shaded area marks the region of incorrect response for the COG method, i.e. $ B^S_{\rm eff} < B^N_{\rm eff}$,
and defines the lower limit for magnetic field detections} 
\label{Fig:14}
\end{minipage}
\end{figure*}
Owing to the linear superposition of the signal and noise contribution in Eq. (\ref{Eq:5.3.1}), 
we do not depend on an underlying magnetic field and only need to simulate, besides the Stokes~$I$ profile, the noise part of the profile.
To quantify the magnetic response to the noise, we calculated the effective longitudinal magnetic field for each synthetic
noise 
profile with the COG and the OMP methods. 
To obtain a statistical meaningful average response for each of 20 noise levels, we computed the 
mean of the absolute effective field from each of the corresponding 10~000 field values. 

In Fig.~\ref{Fig:14}, we plotted the magnetic response of the COG and of the OMP method
over the various noise levels. 
From the response curve, we can immediately identify how much of the noise is falsely interpreted as a magnetic field.
As the noise level increases, the COG method is increasingly affected by the noise, and more and more
of the content of each noise profile is interpreted as an effective longitudinal magnetic field.
For the COG method, a noise level of 10$^{-3}$ results, on average, in an effective longitudinal magnetic field of 101 G.
On the right side of Fig.~\ref{Fig:14}, we expand the region between 10$^{-5}$ to 10$^{-4}$. A few times 10$^{-5}$ 
is the region where many of the recently detected weak magnetic fields are reported 
\citep[e.g.,][]{Auriere09,Ligni09,Auriere10,Grunhut10,Konsta10,Petit11,Fossati13}. Despite the low noise level, 
we see that the COG method interprets a non-negligible amount of the noise as a magnetic field (between 2 and 10 G).

The shaded areas in Fig.~\ref{Fig:14} mark the region of magnetic field values that can no longer be properly
interpreted by the COG method, i.e., the region where $ B^S_{\rm eff} < B^N_{\rm eff}$. 
In that sense the response curves set the lower limit or a noise threshold 
for the estimation of weak magnetic fields. 
However,  even when the true field strength is above these 
threshold values, the \emph{relative} contribution from the noise can severely compromise the field estimation 
of the COG method as was shown in Sect.\ref{Sect:5.2}.
For the magnetic OMP method, on the other hand, we see from Fig.~\ref{Fig:14} that the 
response is largely flat. 

The magnetic field estimation from the OMP method is almost unaffected by the 
noise and shows only a small increase with the noise level.  The incorrectly attributed field values are no higher than 0.7 G. 
This has its origin in the multiresolution thresholding scheme implemented in the OMP algorithm (see Sect.\ref{Sect:3.3}). 

Only those signal or noise features that reach
a certain significance are interpreted and quantified. 
This again results in the good performance and robustness against the contributing noise. 
A dark gray area that shows the incorrect response region of the OMP method is not visible 
in Fig.~\ref{Fig:14} due to the low noise response of the OMP method.
The specific values for the noise response of the COG method
also depend of course on the line parameters, 
such as wavelength, equivalent width, and Land\'e factor. However, the deviations from our results are 
expected to be small because the spread in the derived average parameters for multiline reconstruction 
techniques that use line lists of many hundreds or thousands of lines are relatively small. 
The spectral resolution is also a contributing factor to the error of the COG method since the 
statistical fluctuations that enter into the evaluation of the first-order moment directly depend 
on the number of available wavelength points, such that lower spectral resolutions on average lead to 
larger errors (i.e., to a stronger noise response). 
However, a simulation and critical assessment of all these contributing factors in the COG estimation, 
are not subjects of the current paper. 

\section{Summary}
\label{Sect:6}
This work presents a novel technique for detecting and quantifying stellar magnetic fields.
By employing a sparse profile approximation in terms of an orthogonal matching pursuit 
algorithm, we were able to provide a robust method for detecting and 
estimating the effective longitudinal magnetic field.

By introducing the concept of the apparent longitudinal magnetic field, which 
describes the maximum of the resolvable absolute longitudinal surface field over the 
stellar disk, we provided a complementary measure to the effective or mean 
longitudinal magnetic field. For rapidly rotating stars, the definition of the 
apparent longitudinal magnetic field allows a quantification of the field even 
in situations where a small-scale and balanced surface field would otherwise 
result in a vanishing mean longitudinal magnetic field.

It was shown by an extensive numerical simulation with our \emph{iMap} code \citep{Carroll12}
that the accuracy of the 
OMP method is remarkably good with a mean absolute percentage error of only 1.87 \% 
for the effective longitudinal magnetic field and 4.67 \% for the apparent longitudinal magnetic field.
However, the real strength of the new approach is its robustness against noise. Even for
relative noise levels of unity, i.e., when the variance of the noise has the same magnitude as the 
actual Stokes~$V$ signal, the OMP method has a mean error of only
18 \%, whereas the conventional estimation by the COG method
yields a mean error of almost 50 \%.

In an effort to understand the limitations of the conventional COG method for estimating
weak magnetic fields from faint and noisy Stokes~$V$ signals, we 
simulated noise profiles in a low-noise regime from an absolute noise level of 10$^{-3}$ 
down to 10$^{-5}$. This is a regime that is currently reached by multiline reconstruction techniques 
such as LSD \citep{Donati97} or SVD \citep{Carroll12}
and will be reached by the next generation of spectropolarimeters like
PEPSI at the 2$\times$8.4m Large Binocular Telescope (LBT) \citep{Strassmeier08}. 
In these low-noise regimes, however, a non-negligible fraction of the noise was 
incorrectly interpreted as a magnetic field by the COG method. Depending on the noise level, 
these falsely identified field values are between 2 and 100 G, making any attempt to estimate 
weak fields of a similar strength extremely difficult. The magnetic OMP method, on the other  
hand, again shows a remarkable robustness over the entire range of noise levels, where 
only 0.1 to 0.7 G are incorrectly attributed to a magnetic field. 
This makes the magnetic OMP method the method of choice for estimating weak magnetic fields 
from noisy Stokes profiles.

From a numerical perspective, the algorithm can be easily implemented and  
the iterative process is fast to evaluate.\footnote{An IDL-Code of the magnetic OMP method
is available under http://www.aip.de/People/tcarroll/ and http://www.aip.de/Members/tcarroll/}
The magnetic OMP algorithm provides a viable tool for detecting and quantifying magnetic fields
from spectropolarimetric observations. Although the focus in this work has been placed
on disk-integrated Stokes profiles of stellar magnetic fields, this method
is also directly applicable to resolved solar magnetic field observations. 

\begin{acknowledgements}
The authors would like to thank the anonymous referee for the constructive and helpful
comments, which helped to improve this manuscript.  
\end{acknowledgements}

\end{document}